\pdfoutput=1
\documentclass[11pt,twoside,a4paper,cmspaper,final,collab]{cms-tdr}
\def\svnVersion{eecdbe4}\def\svnDate{2023/07/11}\def\cmsCernNoTag{CERN-EP-2022-010}\def\cmsCernDate{\today}\def\cmsMessage{Published in the European Physical Journal C as \href{http://dx.doi.org/10.1140/epjc/s10052-023-11581-0}{\doi{10.1140/epjc/s10052-023-11581-0}.}}
\begin{document}\cmsNoteHeader{HIG-20-018}

\newlength\cmsFigWidthA
\ifthenelse{\boolean{cms@external}}{\setlength\cmsFigWidthA{0.465\textwidth}}{\setlength\cmsFigWidthA{0.52\textwidth}}
\newlength\cmsFigWidthB
\ifthenelse{\boolean{cms@external}}{\setlength\cmsFigWidthB{0.499\textwidth}}{\setlength\cmsFigWidthB{0.60\textwidth}}
\newlength\cmsFigWidthC
\ifthenelse{\boolean{cms@external}}{\setlength\cmsFigWidthC{0.499\textwidth}}{\setlength\cmsFigWidthC{0.60\textwidth}}
\newlength\cmsFigWidthD
\ifthenelse{\boolean{cms@external}}{\setlength\cmsFigWidthD{0.499\textwidth}}{\setlength\cmsFigWidthD{0.70\textwidth}}
\newlength\cmsFigWidthF
\ifthenelse{\boolean{cms@external}}{\setlength\cmsFigWidthF{0.499\textwidth}}{\setlength\cmsFigWidthF{0.70\textwidth}}
\newlength\cmsFigWidthE
\ifthenelse{\boolean{cms@external}}{\setlength\cmsFigWidthE{0.49\textwidth}}{\setlength\cmsFigWidthE{0.73\textwidth}}

\newcommand{\Wjets}{{\PW}+jets\xspace}
\newcommand{\Zjets}{{\PZ}+jets\xspace}
\newcommand{\Zjet}{{\PZ}+jet\xspace}
\newcommand{\photonjet}{photon+jet\xspace}
\newcommand{\Powheg}{\POWHEG} 
\newcommand{\LSP}{\ensuremath{\PSGc_{\text{S}}^0}\xspace}
\newcommand{\MSUSY}{\ensuremath{m_{\text{SUSY}}}\xspace}
\newcommand{\MLH}{\ensuremath{m_{\PH_1}}\xspace}
\newcommand{\LH}{\ensuremath{\PH_1}\xspace}
\newcommand{\HSM}{\ensuremath{\PH_{\text{SM}}}\xspace}
\newcommand{\DM}{\ensuremath{\Delta_{m}}\xspace}
\newcommand{\RM}{\ensuremath{R_{m}}\xspace}
\newcommand{\MNLSP}{\ensuremath{m_{\PSGczDt}}\xspace}
\newcommand{\MZ}{\ensuremath{m_{\PZ}}\xspace}
\newcommand{\MLSP}{\ensuremath{m_{\LSP}}\xspace}
\newcommand{\CR}{CR\xspace}
\newcommand{\VR}{VR\xspace}
\newcommand{\TR}{TR\xspace}
\newcommand{\SBY}{\ensuremath{\hat{U}_{i}}\xspace}
\newcommand{\SB}{\ensuremath{U_{i}}\xspace}
\newcommand{\SBo}{\ensuremath{U_{1}}\xspace}
\newcommand{\Akin}{\ensuremath{A_{\text{kin}}}\xspace}
\newcommand{\fig}{Fig.}
\newcommand{\figfull}{Figure}
\newcommand{\eq}{Eq.}
\newcommand{\BR}{\ensuremath{\mathcal{B}}\xspace}
\newcolumntype{Y}[1]{D{,}{\,\!\!\,}{#1}}
\newcommand{\Trule}{\rule{0pt}{2.3ex}}

\cmsNoteHeader{HIG-20-018}
\title{Search for light Higgs bosons from supersymmetric cascade decays in \texorpdfstring{$\Pp\Pp$}{pp} collisions at \texorpdfstring{$\sqrt{s}=13\TeV$}{sqt(s)=13 TeV}}

\date{\today}

\abstract{
A search is reported for pairs of  light Higgs bosons (\LH) produced in supersymmetric cascade decays  in final states with small missing transverse momentum. A data set of LHC $\Pp\Pp$ collisions collected with the CMS detector at $\sqrt{s}=13\TeV$ and corresponding to an integrated luminosity of 138\fbinv is used. The search targets events where both \LH bosons decay into $\bbbar$ pairs that are reconstructed as large-radius jets using substructure techniques. No evidence is found for an excess of events beyond the background expectations of the standard model (SM). Results from the search are interpreted in the next-to-minimal supersymmetric extension of the SM, where a ``singlino'' of small mass leads to squark and gluino cascade decays that can predominantly end in a highly Lorentz-boosted singlet-like \LH and a singlino-like neutralino of small transverse momentum. Upper limits are set on the product of the squark or gluino pair production cross section and the square of the $\bbbar$ branching fraction of the \LH in a benchmark model containing almost mass-degenerate gluinos and light-flavour squarks. Under the assumption of an SM-like $\LH\to\bbbar$ branching fraction, \LH bosons with masses in the range 40--120\GeV arising from the decays of squarks or gluinos with a mass of 1200 to 2500\GeV are excluded at 95\% confidence level.
}

\hypersetup{%
pdfauthor={CMS Collaboration},%
pdftitle={Search for light Higgs bosons from supersymmetric cascade decays in pp collisions at sqrt(s) = 13 TeV},%
pdfsubject={CMS},%
pdfkeywords={CMS, Higgs, supersymmetry, NMSSM, large-radius jets, double-b tagging, soft drop}}

\maketitle

\section{Introduction}\label{sec_introduction}

\begin{figure}[hbtp]
    \centering
    \includegraphics[width=\cmsFigWidthA]{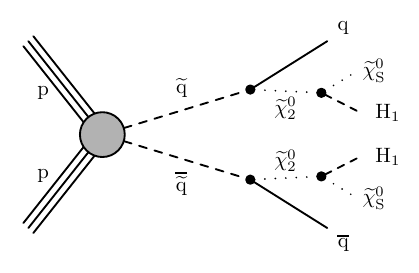}
    \caption{
    Diagram of squark pair production and subsequent cascade decay in the benchmark signal model.
    The particle \PSGczDt is the next-to-LSP,
    \LSP is the singlino-like LSP,
    and \LH is the $CP$-even singlet-like Higgs boson.
    }
    \label{fig:signalFeynmanDiagram}
    \end{figure}

{\tolerance=500
This paper presents a search for pairs of
light Higgs bosons (\LH) produced in supersymmetric (SUSY)~\cite{Ramond:1971gb,Golfand:1971iw,Neveu:1971rx,Volkov:1972jx,Wess:1973kz,Wess:1974tw,Fayet:1974pd,Nilles:1983ge} cascade decays
in final states with small missing transverse momentum (\ptmiss).
Such events can arise from
the pair production of squarks (\PSq) and gluinos (\PSg) in the next-to-minimal supersymmetric extension of the standard model (SM)~\cite{Ellwanger_2010}
when the lightest SUSY particle (LSP) is a singlino-like neutralino (\LSP) of small mass~\cite{Ellwangera:modelMotivationLightHiggs}.
The \LSP mass eigenstate is dominated by the singlino component and has only small couplings to other SUSY particles,
suppressing direct squark or gluino decays to the \LSP.
Squarks and gluinos decay via the next-to-LSP \PSGczDt into a \LSP and a Higgs, \PZ, or \PW~boson~\cite{Ellwangera:modelMotivationLightHiggs,Ellwangera:modelMotivationSMHiggs}.
The case of a singlet-like $CP$-even \LH, shown in \fig~\ref{fig:signalFeynmanDiagram}, is the focus of this search.
When the \LSP has a far smaller mass than the \LH and the phase space for the decay $\PSGczDt\to \LH+\LSP$ is small,
the \LH carries much larger momentum than the \LSP.
In such \ptmiss-suppressed scenarios, the key signature for the pair production of squarks and gluinos
is a pair of Lorentz-boosted \LH bosons.
\par}

This search targets events with two highly boosted \LH bosons that decay into $\bbbar$ pairs that are reconstructed as large-radius jets using substructure techniques.
This is the first search at the LHC to focus on this type of event, where particles invisible to the detector have only small transverse momentum (\pt) and therefore the events are not selected by searches requiring significant \ptmiss~\cite{Titterton:2018pba, Ellwangera:modelMotivationLightHiggs}.
Previous searches by the ATLAS and CMS experiments with similar final states have considered events with two boosted SM Higgs bosons and large values of \ptmiss~\cite{Sirunyan:2017bsh,SUS-20-004},
or two SM Higgs bosons in resolved final states where each of the four \Pb quarks is reconstructed as a separate jet, with either small~\cite{Aaboud:2018htj} or large~\cite{Khachatryan:2014mma,Sirunyan:2017obz,Aaboud:2018htj,SUS-20-004} values of \ptmiss.
This search uses data from $\Pp\Pp$ collisions
collected by the CMS detector at $\sqrt{s}=13\TeV$ during 2016--2018,
corresponding to an integrated luminosity of 138\fbinv~\cite{CMS-LUM-17-003,lumi2017,lumi2018}.

\section{Benchmark signal model}\label{sec_signalModelAndMotivation}

{\tolerance=500
A benchmark signal model is established following the work of Ellwanger and Teixeira~\cite{Ellwangera:modelMotivationLightHiggs,Ellwangera:modelMotivationSMHiggs}.
The eight first- and second-generation squarks are assumed mass-degenerate at the mass \MSUSY,
and the gluino mass is set at 1\% larger.
The small gluino-squark mass gap means that
the kinematics of the final-state particles are very similar
in the $\PSq\PSq$, $\PSq\PSg$, and $\PSg\PSg$ production modes,
as little momentum is transferred to the quark in the $\PSg\to\PSq+\cPq$ decay.
All SUSY particles other than gluinos and those shown in \fig~\ref{fig:signalFeynmanDiagram} are assumed decoupled.
\par}
{\tolerance=2400
This search targets squarks and gluinos with $\MSUSY > 1200 \GeV$.
Less massive squarks and gluinos can be probed by \ptmiss-based searches, owing to their larger pair-production cross sections~\cite{Titterton:2018pba}.
Smaller \MSUSY values can also lead to smaller \pt of the \LH 
than is necessary for the $\bbbar$ pair to be merged in a single jet.
The cross sections ($\sigma$) for the signal probed in this search, calculated 
at next-to-leading order (NLO) accuracy in the strong coupling constant (\alpS) including approximate next-to-NLO (NNLO) corrections
and next-to-next-to-leading logarithmic (NNLL) soft gluon corrections~\cite{bib-nlo-nll-01,bib-nlo-nll-02,bib-nlo-nll-03,bib-nlo-nll-04,bib-nlo-nll-05,bib-nnll-02,bib-nnll-03,bib-nnll-04,bib-nnll},
are shown in Table~\ref{table:sigSamples_MSUSY}.
\par}

\begin{table}[hbtp]
    \centering
    \topcaption{
    Inclusive pair-production cross sections
    calculated at approximately NNLO and NNLL in \alpS~\cite{bib-nlo-nll-01,bib-nlo-nll-02,bib-nlo-nll-03,bib-nlo-nll-04,bib-nlo-nll-05,bib-nnll-02,bib-nnll-03,bib-nnll-04,bib-nnll}
    for squark mass \MSUSY and gluino mass 1\% larger.
    The quoted uncertainty is obtained from variations in the choice of scales, parton distribution functions, and \alpS.
    }
    \begin{tabular}{c|Y{4.4}|Y{8.2}}
        \MSUSY [{\GeVns}] & \multicolumn{1}{c|}{$\sigma(\Pp\Pp \to \PSq\PSq,\PSq\PSg,\PSg\PSg)$ [fb]} & \multicolumn{1}{c}{Uncertainty} \\ [0.3ex]
        \hline
        \Trule
        1200	& 580,	& $8\%$ \\
        1600	& 69,	& $9\%$ \\
        2000	& 10,	& $11\%$ \\
        2200	& 4,.1	& $13\%$ \\
        2400	& 1,.6	& $14\%$ \\
        2600	& 0,.67	& $16\%$ \\
        2800	& 0,.27	& $18\%$ \\
        \end{tabular}
        \label{table:sigSamples_MSUSY}
        \end{table}

The values considered of the \LH mass (\MLH)
and the corresponding $\LH\to\bbbar$ branching fractions (\BR)
are shown in Table~\ref{table:sigSamples_MH}.
Only events where both \LH bosons decay into $\bbbar$ pairs are used as signal.
The \BR values are chosen to
be those of an SM-like Higgs boson (\HSM) of the corresponding mass~\cite{Ellwangera:modelMotivationLightHiggs}, 
as calculated using \HDECAY~6.61~\cite{DJOUADI199856, DJOUADI2019214}.
The \BR values decrease for larger \LH masses as the virtual $\PW\PW^{*}$ and $\PZ\PZ^{*}$ decay channels, both of which have sizeable leptonic branching fractions, become more accessible.
The region $\MLH < \MZ$
is therefore where the \ptmiss-suppressed all-jet signature is of greatest experimental importance.
Nevertheless, to preserve generality, this search attempts to probe as much of the region $\MLH < 125\GeV$ as possible.

\begin{table*}[hbtp]
    \centering
    \topcaption{
    The \MLH values in this search
    and corresponding $\LH\to\bbbar$ branching fractions.
    }
    \begin{tabular}{c|cccccccccccc}
      \MLH [{\GeVns}] & 30 & 35 & 40 & 50 & 60 & 70 & 80 & 90 & 100 & 110 & 120 & 125 \\ [0.3ex]
      \hline
      \Trule
      $\BR(\LH\to\bbbar)$ & 0.86 & 0.86 & 0.86 & 0.86 & 0.85 & 0.84 & 0.83 & 0.81 & 0.79 & 0.75 & 0.65 & 0.58 \\
    \end{tabular}
    \label{table:sigSamples_MH}
    \end{table*}

In addition to \MLH and \MSUSY, there are two other unknown masses in the benchmark model:
those of the \LSP and the \PSGczDt.
The corresponding degrees of freedom are parameterised by
$\RM \equiv \MLH / \MNLSP$ and
$\DM \equiv \MNLSP - \MLH - \MLSP$.
The \ptmiss-suppressed signature arises for values of \RM close to unity, provided $\DM>0$ to permit the $\PSGczDt\to \LH+\LSP$ decay.
In this case, the phase space for the \PSGczDt decay is small and the \LSP has much smaller mass than the \LH,
so the \LSP always carries much less momentum than the \LH.
The \ptmiss-suppressed signature probed in this search is representative
of a significant part of the model parameter space
since the momenta of reconstructed objects do not exhibit a strong dependence on \RM and \DM in the region $\RM>0.9$.
Models with smaller \RM can be probed by \ptmiss-based searches~\cite{Titterton:2018pba, Ellwangera:modelMotivationLightHiggs}.
For the benchmark model, the values $\RM = 0.99$ and $\DM = 0.1\GeV$ are assumed.

Branching fractions of unity are assumed for
the decays $\PSq\to\cPq+\PSGczDt$
and $\PSGczDt\to \LH+\LSP$.
In the \RM and \DM region of the benchmark model,
this is true
except where $\MNLSP > \MZ + \MLSP$.
In that case, the $\PSGczDt \to \PZ+\LSP$ decay is permitted
if the \PSGczDt has a higgsino component~\cite{Ellwangera:modelMotivationSMHiggs}.
However, the \PSGczDt is expected to be mainly bino-like for relevant values of its mass~\cite{Ellwangera:modelMotivationLightHiggs}.
For configurations where the \LH mass is close to that of the \HSM, the decay $\PSGczDt\to \HSM+\LSP$ is also possible.
The signatures for such \LH and \HSM bosons are indistinguishable in this search.
The assumption that the branching fraction for $\PSGczDt\to \LH+\LSP$ decay is 100\% can therefore be relaxed to the assumption that the branching fractions to \LH and \HSM sum to unity.

\section{The CMS detector}\label{sec_detector}

{\tolerance=500
The central feature of the CMS apparatus is a superconducting solenoid of 6\unit{m} internal diameter,
providing a magnetic field of 3.8\unit{T}.
A silicon pixel and strip tracker,
a lead tungstate crystal electromagnetic calorimeter,
and a brass and scintillator hadron calorimeter,
each composed of a barrel and two endcap sections,
reside within the solenoid volume.
Forward calorimeters extend the pseudorapidity ($\eta$) coverage provided by the barrel and endcap detectors.
Muons are measured in gas-ionisation detectors embedded in the steel flux-return yoke outside the solenoid.
Events of interest are selected using a two-tiered trigger system. The first level, composed of custom hardware processors, uses information from the calorimeters and muon detectors to select events at a rate of around 100\unit{kHz} within a fixed latency of about 4\mus~\cite{Sirunyan:2020zal}. The second level, known as the high-level trigger, consists of a farm of processors running a version of the full event reconstruction software optimised for fast processing, and reduces the event rate to around 1\unit{kHz} before data storage~\cite{Khachatryan:2016bia}.
A more detailed description of the CMS detector,
together with a definition of the coordinate system and the kinematic variables,
can be found in Ref.~\cite{Chatrchyan:2008zzk}.
\par}

\section{Event simulation}\label{sec_eventSamples}

The primary background in this search originates from
multijet production.
Simulated multijet events are used to validate the multijet background estimation based on data (described in Section~\ref{sec_qcdEstimation}),
but are not used for any of the final predictions.
The remaining significant background is from events with vector bosons that decay into quark-antiquark pairs.
Simulated events are used to determine the contributions from \ttbar, \Zjets, and \Wjets production.
The expected yields from all other SM sources of background are found to be negligible.

{\tolerance=5000
The multijet, \Zjets, and \Wjets processes are simulated at leading order (LO) in perturbative quantum chromodynamics (QCD) using \MGvATNLO~2.4.2~\cite{mg5amcnlo} with up to four additional partons at the matrix element (ME) level.
Simulated signal events for each pair of \MLH and \MSUSY values of the benchmark model
are generated at LO at the ME level with up to one additional parton using \MGvATNLO~2.3.3.
The MLM~\cite{Alwall:2007fs} prescription is used to match partons from the LO ME calculations to those from the parton showers.
Simulated \ttbar\ events
are produced
at NLO in QCD at the ME level with the \Powheg\ $\text{v}2.0$~\cite{Frixione:2007nw,bib:powheg0,bib:powheg,bib:powheg2} generator.
The NNPDF2.3, NNPDF3.0, and NNPDF3.1~\cite{bib:NNPDF,Ball2015,nnpdf23bib,nnpdf3bib} parton distribution functions (PDFs) are used for the signal, 2016 background, and 2017--2018 background simulations, respectively.
The parton shower and hadronisation are performed via \PYTHIA~8.2~\cite{Sjostrand:2007gs}.
The CUETP8M1~\cite{bib:CUETP8tune,Skands:2014pea} tune is used for the signal and 2016 background simulations, while the CP5 tune~\cite{CP15bib} is used for the 2017 and 2018 background simulations.
The cross section used to normalise the \ttbar simulation is calculated at NNLO+NNLL in QCD~\cite{Czakon:2011xx}, and those for \Zjets and \Wjets are calculated at NNLO in QCD~\cite{Gavin:2010az,Gavin:2012sy,Li:2012wna}.
Additional $\Pp\Pp$ interactions within the same or nearby bunch crossings (pileup) are simulated for all events
according to the distribution of the number of interactions observed in each bunch crossing~\cite{Sirunyan:2018nqx}.
The interactions of particles with the CMS detector are simulated using \GEANTfour~\cite{bib:geant}.
\par}

\section{Object reconstruction and event selection}\label{sec_eventSelection}

{\tolerance=500
The data are collected using triggers based on
the scalar sum of jet \pt (\HT),
with a requirement of $\HT>900\GeV$ (2016) and $\HT>1050\GeV$ (2017 and 2018).
Events are reconstructed offline using a particle-flow (PF) algorithm~\cite{bib:Sirunyan:2017ulk}
that reconstructs and identifies each individual particle (PF candidate) in an event using an optimised combination of information from the components of the CMS detector.

Jets are reconstructed by clustering the PF candidates using the anti-\kt clustering algorithm~\cite{bib:antikt}, as implemented in the \FASTJET package~\cite{Cacciari:2011ma}.
A distance parameter of 0.4 or 0.8 is used for standard- and large-radius jets, referred to as AK4 and AK8 jets, respectively.
The jet momentum is defined as the vectorial sum of all particle momenta in the jet.
To mitigate the effect of pileup, constituent charged PF candidates identified to be originating from vertices other than the primary $\Pp\Pp$ interaction vertex are not used in the clustering algorithm.
The primary vertex is taken to be the vertex corresponding to the hardest scattering in the event, evaluated using tracking information alone, as described in Section 9.4 of Ref.~\cite{CMS-TDR-15-02}.
For AK4 jets, an offset correction is applied to correct for remaining pileup contributions.
For AK8 jets, 
the pileup-per-particle identification algorithm~\cite{Sirunyan:2020foa,Bertolini:2014bba} is used to rescale the momenta of constituent neutral particles
according to the probability they originated from the primary vertex. This probability is based on
a local shape variable that distinguishes between collinear and soft diffuse distributions of the surrounding charged particles that are compatible with the primary vertex.
For all jets, jet energy corrections are derived from simulation to bring the measured average response of jets to that of particle-level jets. In situ measurements of the momentum balance in dijet, \photonjet, \Zjet, and multijet events are used to account for any residual differences in jet energy scale and resolution between data and simulation~\cite{Khachatryan:2016kdb,bib:JER}.
Additional criteria are imposed to reject jets from 
spurious sources, such as electronics noise and detector malfunctions~\cite{CMS-PAS-JME-10-003,CMS-PAS-JME-16-003}.
\par}

The identification of AK8 jets originating from two collimated \PQb quarks (double-\PQb tagging) is integral to the reconstruction of the \LH.
A discriminant is calculated for each jet using a double-\PQb tagging algorithm that combines tracking and vertexing information in a multivariate approach with no strong dependence on jet mass or \pt~\cite{BTV-16-002}.

\begin{figure}[hbtp]
\centering
\includegraphics[width=\cmsFigWidthD]{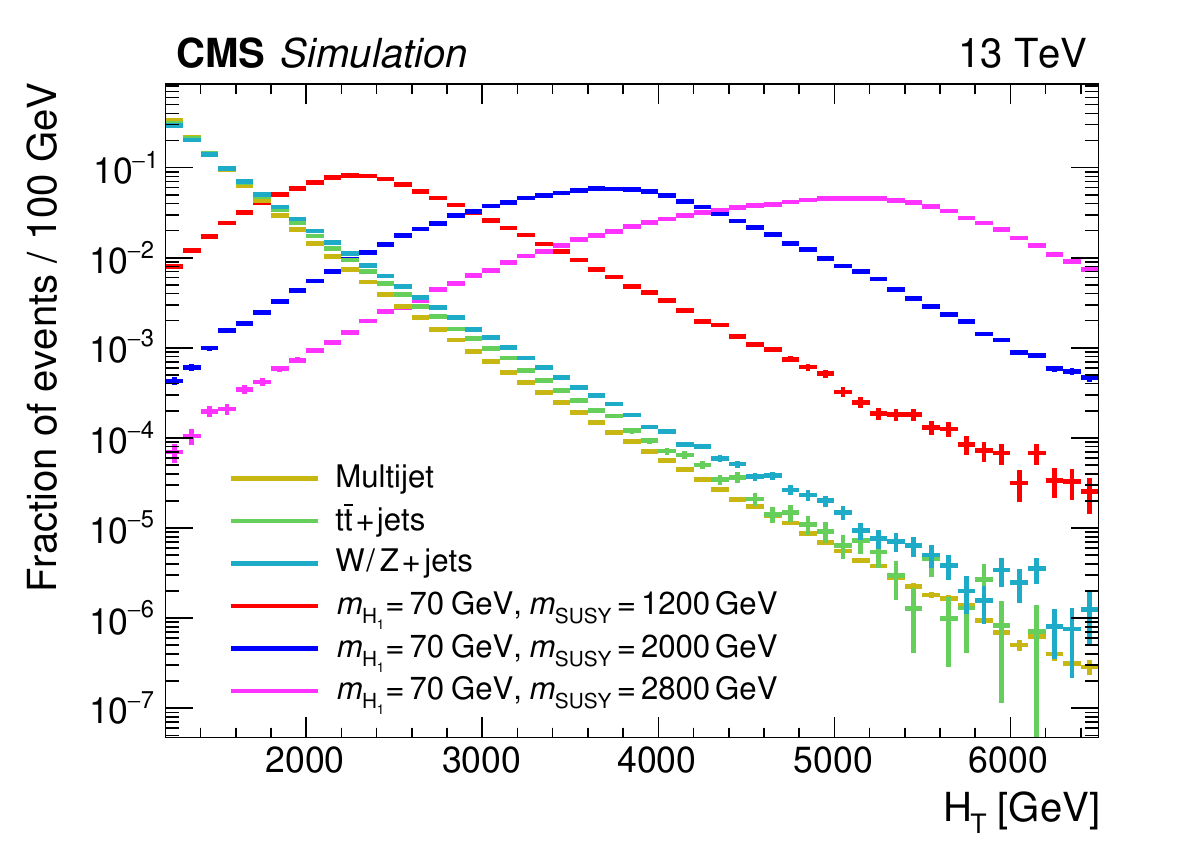}
\caption{
The \HT distribution in signal events with different values of \MSUSY, and in the simulated SM backgrounds, normalised to unit area.
The uncertainties are statistical.
All events satisfy the preselection.
}
\label{fig:kinematicDists}
\end{figure}

The event preselection requires two AK8 jets with $\pt > 170\GeV$
and $|\eta| < 2.4$ (so that they are within the acceptance of the tracker).
If there are more than two candidate AK8 jets, the two with the largest double-\PQb~tag discriminants are selected
as most likely to have originated from $\LH\to\bbbar$ decays.
For the offline analysis, \HT is defined as the scalar \pt sum of
all AK4 jets with $\pt > 40\GeV$ and $|\eta| < 3.0$, including AK4 jets with PF candidates clustered into AK8 jets.
The \HT distributions for various simulated signal and background processes are shown in \fig~\ref{fig:kinematicDists}, after implementing all preselection requirements.
Since the final state contains only jets,
the average signal event \HT depends significantly on \MSUSY, and signal events with $\MSUSY > 1200 \GeV$
tend to have $\HT>1500\GeV$.

Additional requirements based on the expected kinematic properties of signal events are applied after the preselection.
They define the kinematic event selection:
\begin{enumerate}
\item
Both selected AK8 jets must have $\pt > 300\GeV$ and $|\eta| < 2.4$, characteristic of the jets originating from $\LH\to\bbbar$ decay in signal events.
\item
There must be at least one AK4 jet with $\pt > 300\GeV$ and $|\eta| < 3.0$, characteristic of the quarks from squark decays in signal events.
Such jets must be separated by $\Delta R \equiv \sqrt{\smash[b]{(\Delta \phi)^2 + (\Delta \eta)^2}} > 1.4$ from both selected AK8 jets,
to avoid being constructed from the same PF candidates.
\item
The event \HT must exceed 1500\GeV.
\end{enumerate}

Although the offline \HT resolution is better than that of the trigger-level variable,
the offline \HT threshold
is comfortably above the trigger-level \HT requirements.
The trigger efficiency for this analysis is measured using events collected with a single muon trigger with a muon \pt threshold between 24 and 27\GeV. The efficiency for each data-taking year is nearly 100\%.
For the 2018 data, the $|\eta|$ selection for the AK4 jets is reduced from 3.0 to 2.4 to avoid
a region of the endcap electromagnetic calorimeters affected by large losses in crystal transparency,
and therefore increased energy-equivalent electronics noise~\cite{Adams:2255803}.
This change has a negligible effect on signal acceptance for all considered masses.

The fraction of signal events that satisfy the kinematic selection is essentially independent of \MLH.
It increases from about 60 to 80\% as \MSUSY increases from 1200 to 2000\GeV, after which it remains approximately constant.

\subsection{\texorpdfstring{Double-\PQb tag based event selection}{Double-b tag based event selection}}\label{sec_doubleBTagEventSelection}

\begin{figure}[ht]
\centering
\includegraphics[width=\cmsFigWidthB]{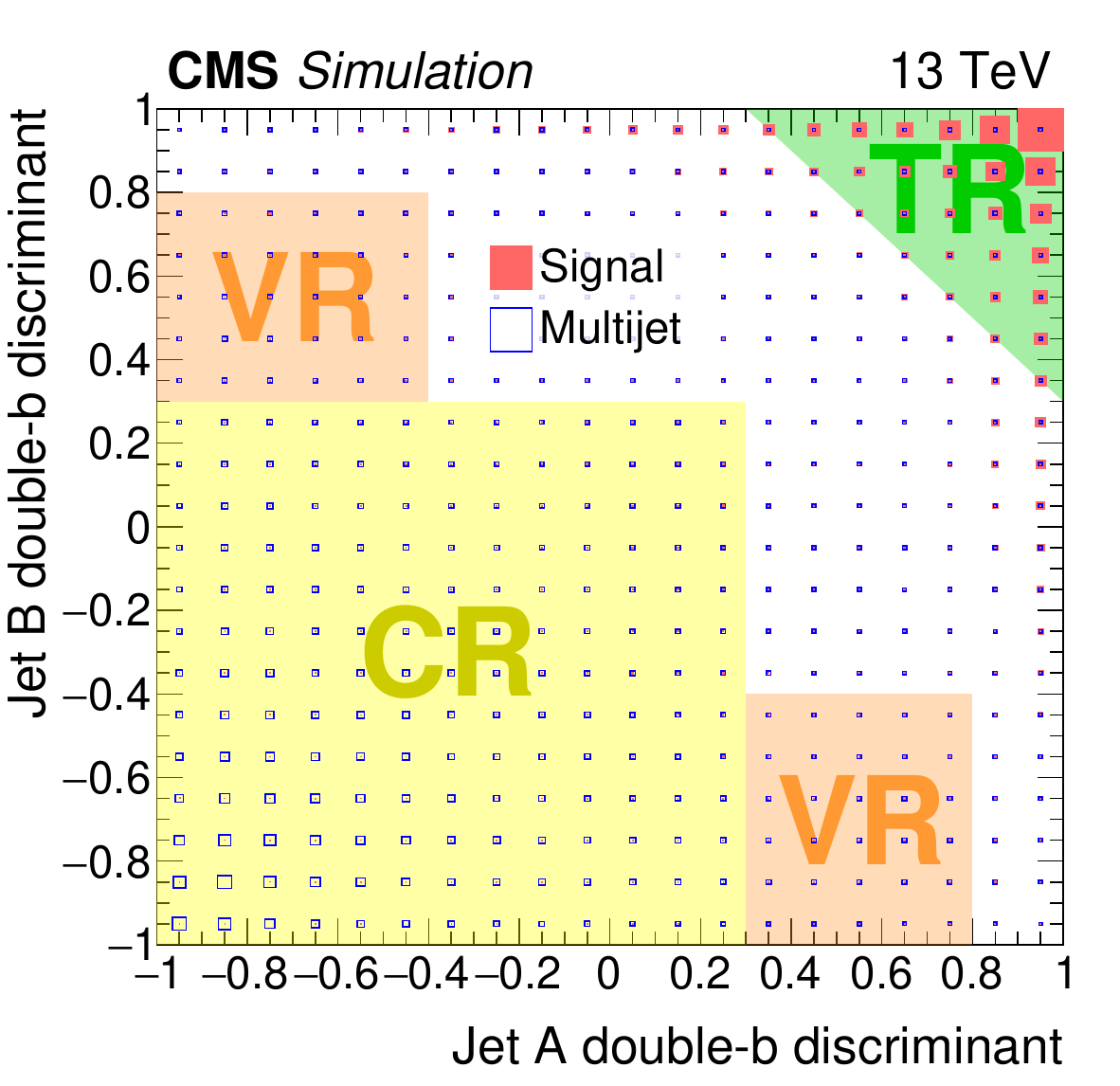}
\caption{
Distributions of simulated signal and multijet events
in the 2D double-\PQb~tag discriminant plane,
where the fractions of events in each bin are represented by the areas of the filled red and open blue squares, respectively.
The signal parameters are $\MLH = 70\GeV$ and $\MSUSY = 2000\GeV$.
The kinematic selection is implemented
with the
masses of the two AK8 jets required to be within the set of signal and sideband mass regions defined in
Section~\ref{sec_softDropMassEventSelection}.
The green, yellow, and orange shaded areas represent the tag region (\TR),
control region (\CR), and validation region (\VR), respectively.
Of the plotted signal events, 65\% fall within the TR.
}
\label{fig:dbtTagAntiControl}
\end{figure}

The two AK8 jets 
that are classified as the $\LH\to\bbbar$ candidates in each event 
are randomly assigned the labels ``A'' and ``B''.
Their double-\PQb~tag discriminants
define a two-dimensional (2D) parameter space,
shown with simulated signal and multijet event distributions in \fig~\ref{fig:dbtTagAntiControl}.
The signal events are expected to contain two $\LH\to\bbbar$ decays and therefore accumulate in the region where both double-\PQb~tag discriminants are large.
The signal-enhanced tag region (\TR) is defined
as the region where the sum of the two double-\PQb~tag discriminants exceeds 1.3, illustrated by
the shaded triangle in \fig~\ref{fig:dbtTagAntiControl}.
Two additional regions are defined in \fig~\ref{fig:dbtTagAntiControl} for use in the multijet background estimation and validation:
the control region (\CR), a multijet-dominated region with negligible signal;
and the validation region (\VR), a more signal-like region where one of the two jets has a large double-\PQb~tag discriminant. The VR is defined sufficiently far from the \TR for the signal contamination to be negligible.

About 50\% of the signal events that satisfy the kinematic selection populate the \TR,
with variation at the level of $\pm10\%$ across 
the \MLH and \MSUSY parameter space considered.
Since the multijet background is dominated by light-flavour quark and gluon initiated jets,
only about 3\% of these events populate the \TR.
For the \ttbar, \Zjets, and \Wjets backgrounds, the corresponding figures are 13, 6, and 3\%, respectively.

\subsection{Soft-drop mass based signal and sideband regions}\label{sec_softDropMassEventSelection}

In signal events, both selected AK8 jets are likely to originate from $\LH\to\bbbar$ decays and therefore have a jet mass close to \MLH.
The multijet background has no resonant mass peak, while the other backgrounds are only expected to exhibit peaks near the known top quark and vector bosons masses,
which means that an accurate reconstruction of the jet mass is important in distinguishing signal from background.
The AK8 jet masses are evaluated
using the ``soft-drop'' algorithm~\cite{Larkoski:2014wba} (with a soft-drop threshold of $z_{\textrm{cut}} = 0.1$ and angular exponent of $\beta = 0$), in which wide-angle soft radiation is removed recursively from a jet.
In signal events this algorithm achieves a relative jet mass resolution from 10\% for $\MLH=125\GeV$ to 20\% for $\MLH=30\GeV$.

\begin{figure}[hbtp]
\centering
\includegraphics[width=\cmsFigWidthB]{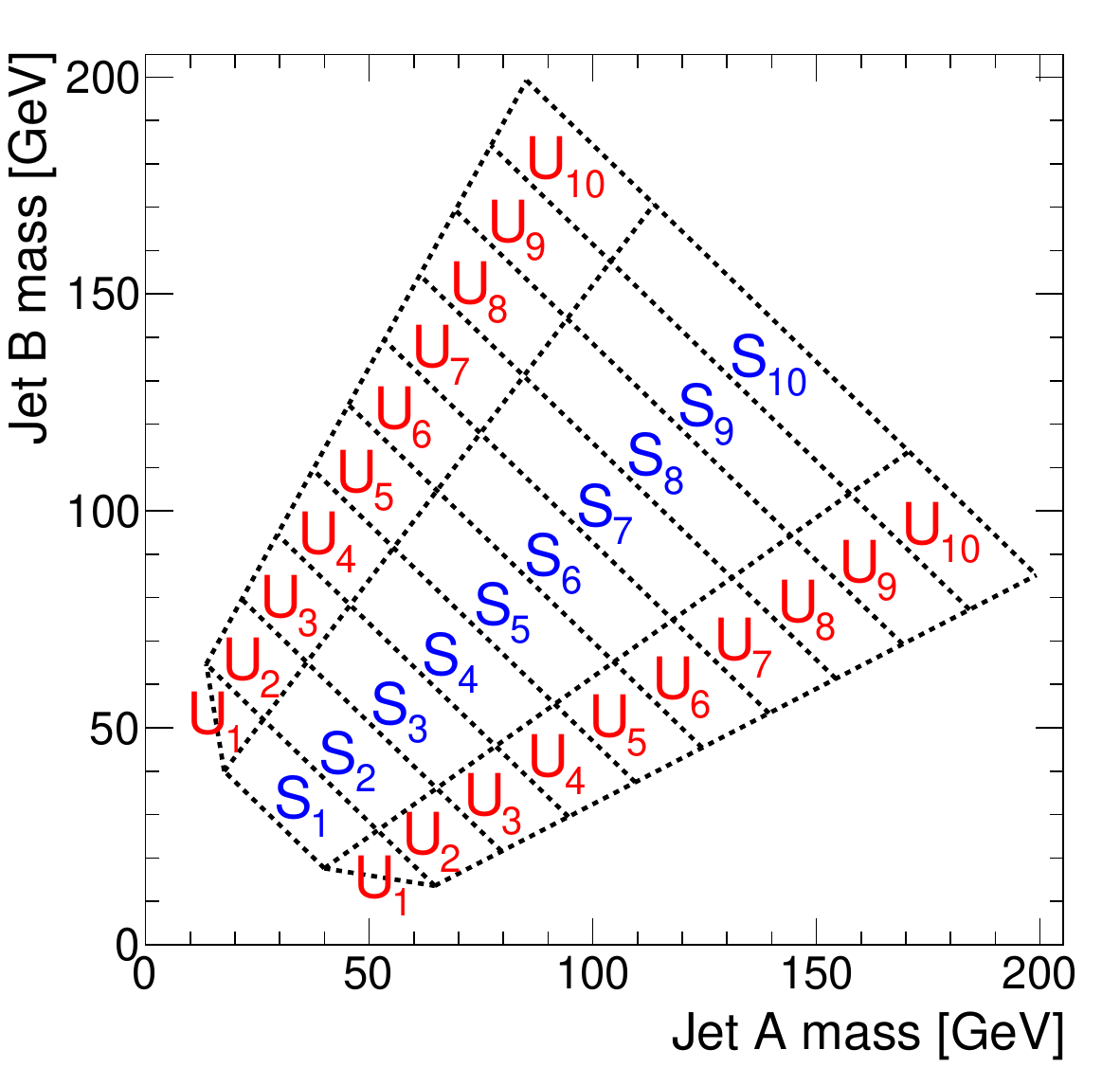}
\caption{
Map of mass regions used in the 2D soft-drop mass plane.
The regions labelled $S_i$ are the signal mass regions,
and the disjoint regions \SB form the corresponding sidebands.
}
\label{fig:massGridEmpty}
\end{figure}

The soft-drop masses of the two AK8 jets define a 2D parameter space, shown in
\fig~\ref{fig:massGridEmpty}, in which 10 signal regions ($S_i$) and 10 sideband regions (\SB) are defined.
The $S_i$ contain events in which the two \LH-candidate jets have approximately the same soft-drop mass.
The width of each $S_i$
corresponds to about four times the experimental soft-drop mass resolution for the relevant simulated value of \MLH.

\begin{figure*}[hbtp]
\centering
\includegraphics[width=0.494\textwidth]{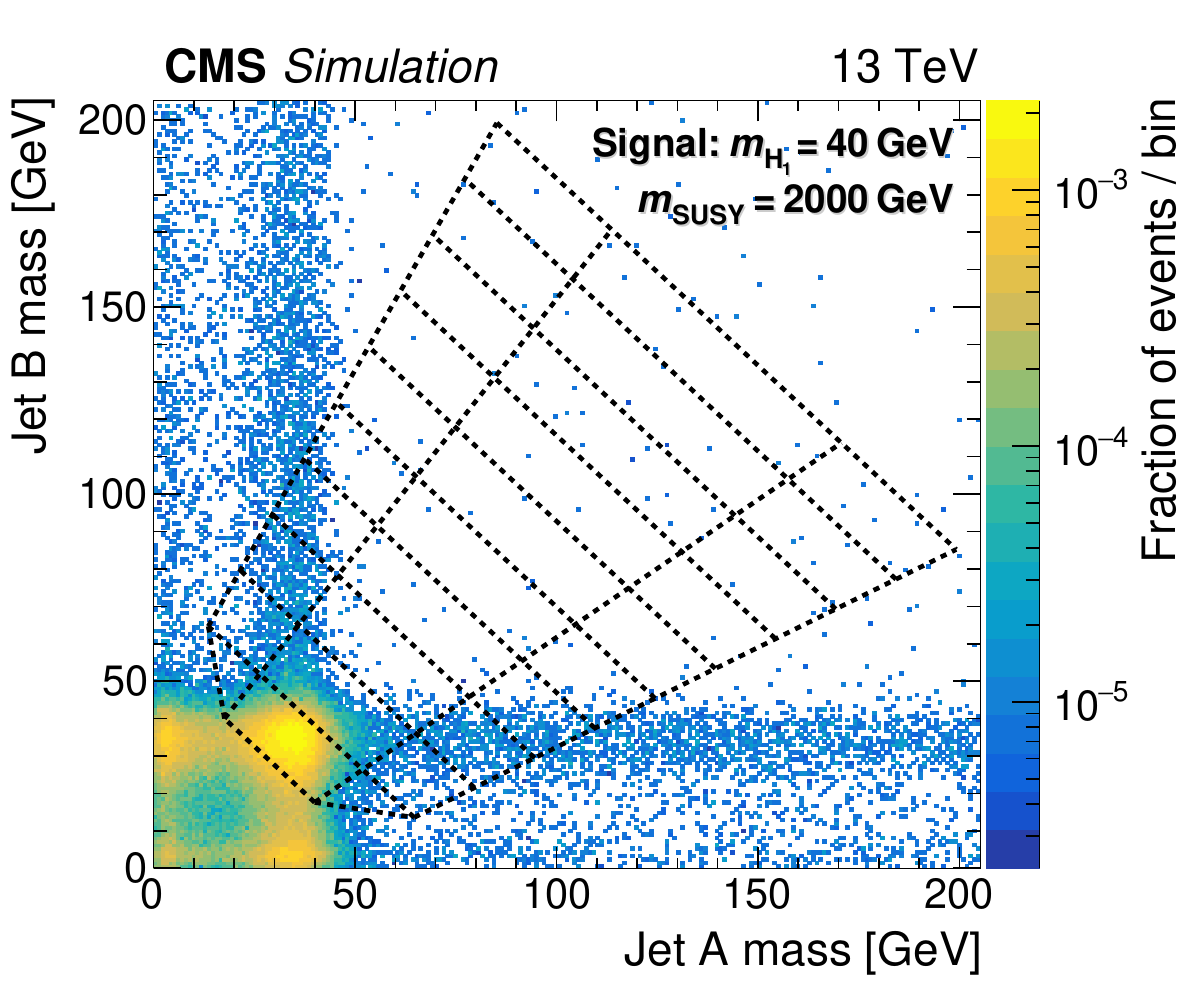}
\includegraphics[width=0.494\textwidth]{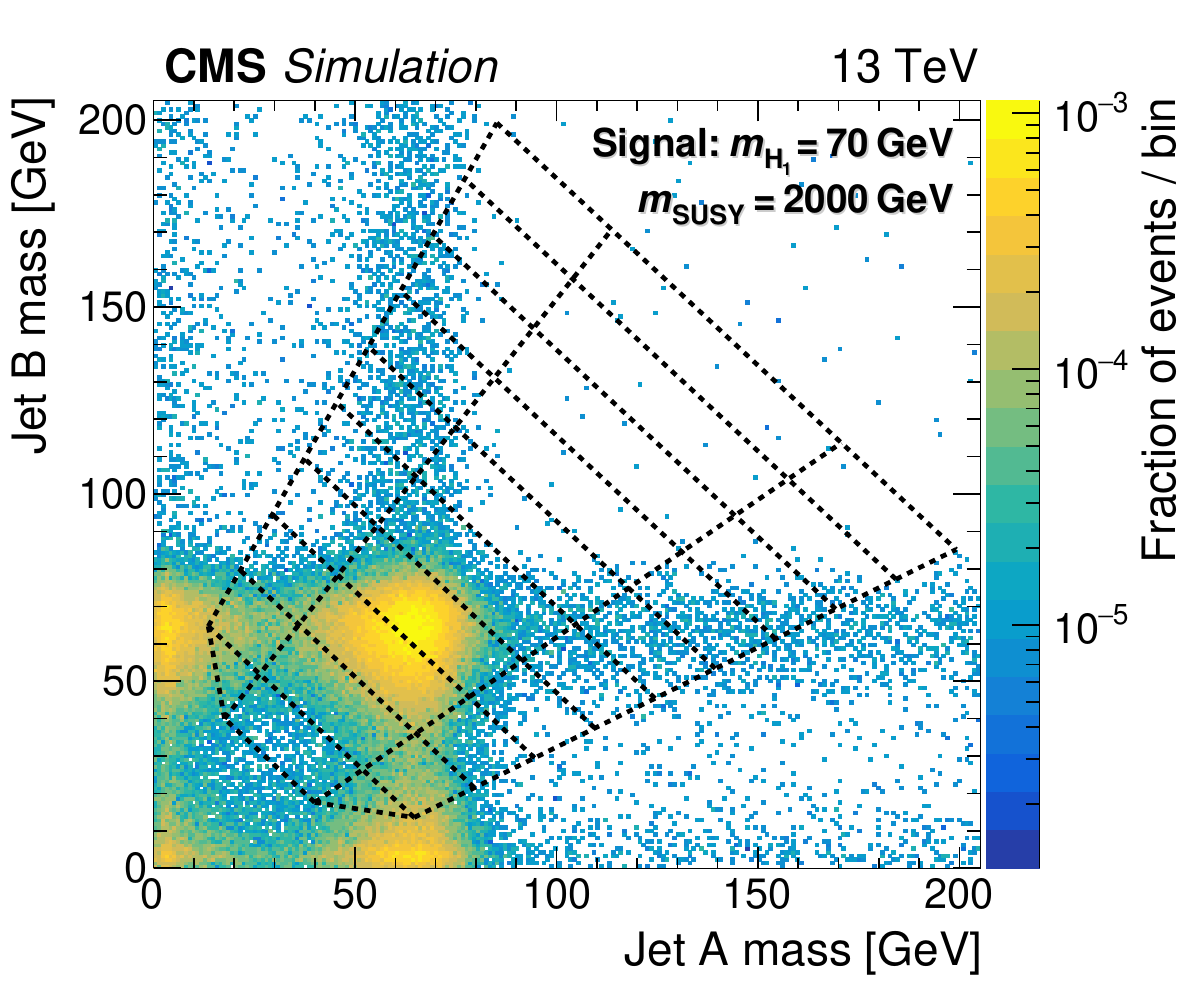}
\includegraphics[width=0.494\textwidth]{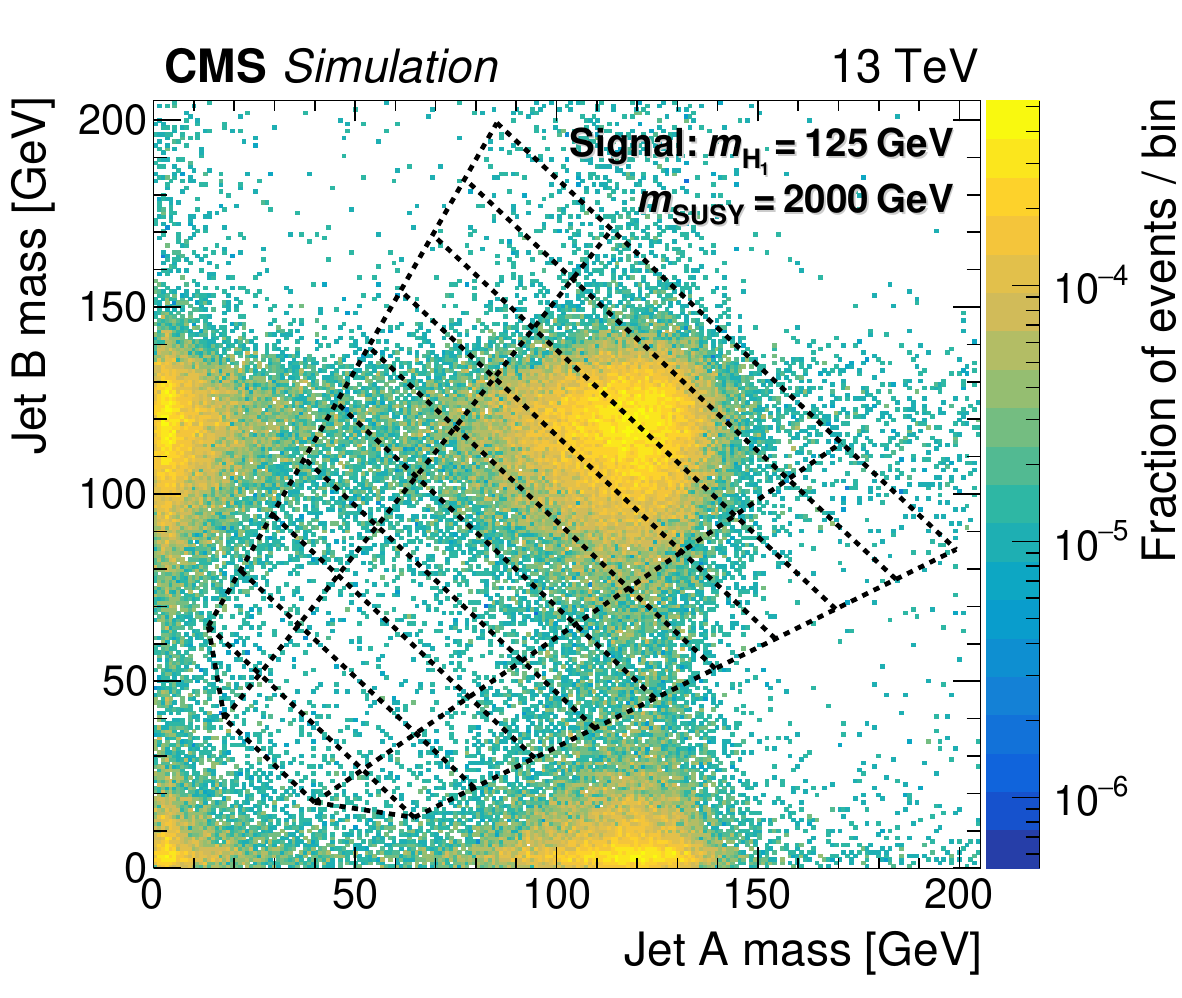}
\includegraphics[width=0.494\textwidth]{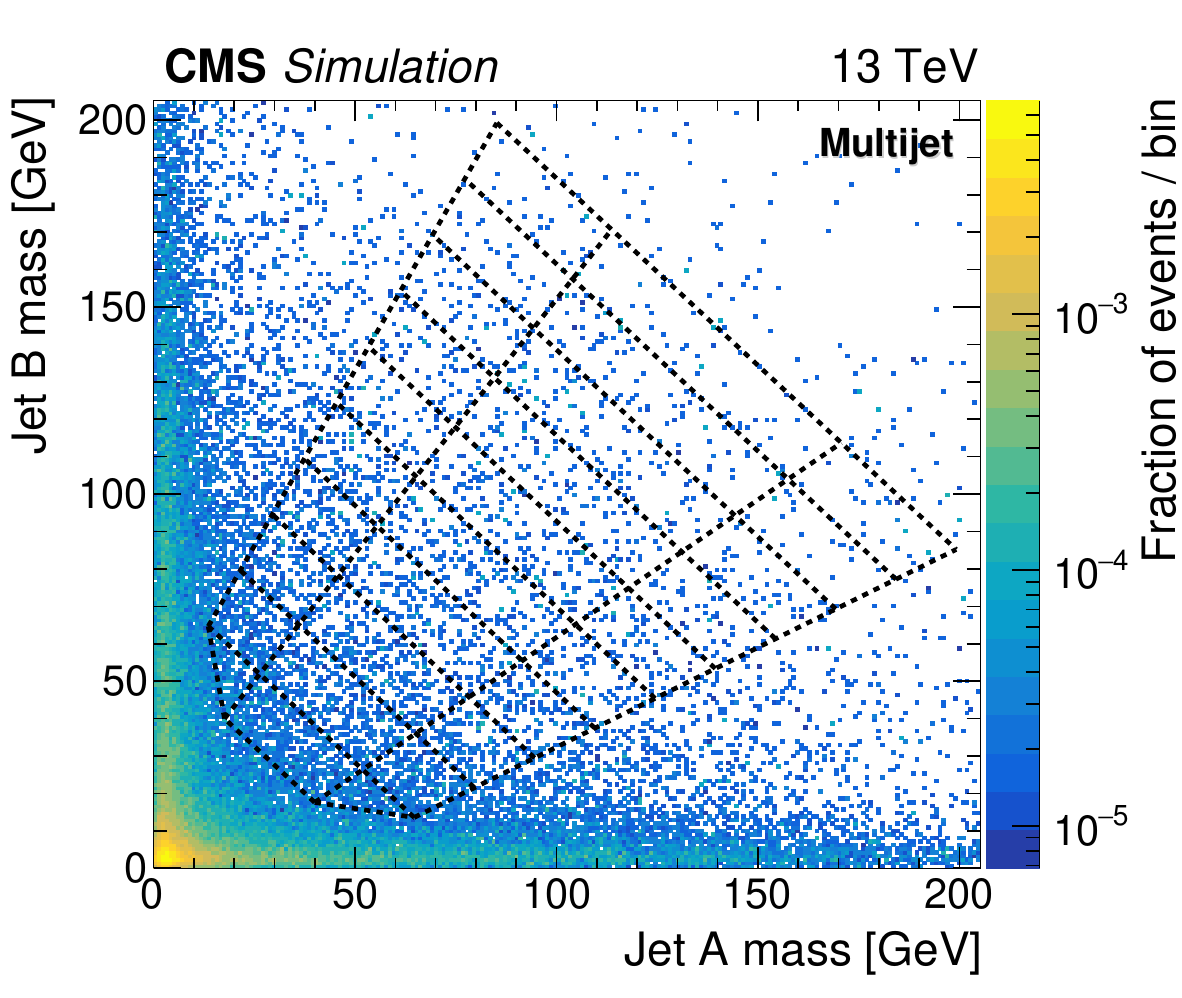}
\includegraphics[width=0.494\textwidth]{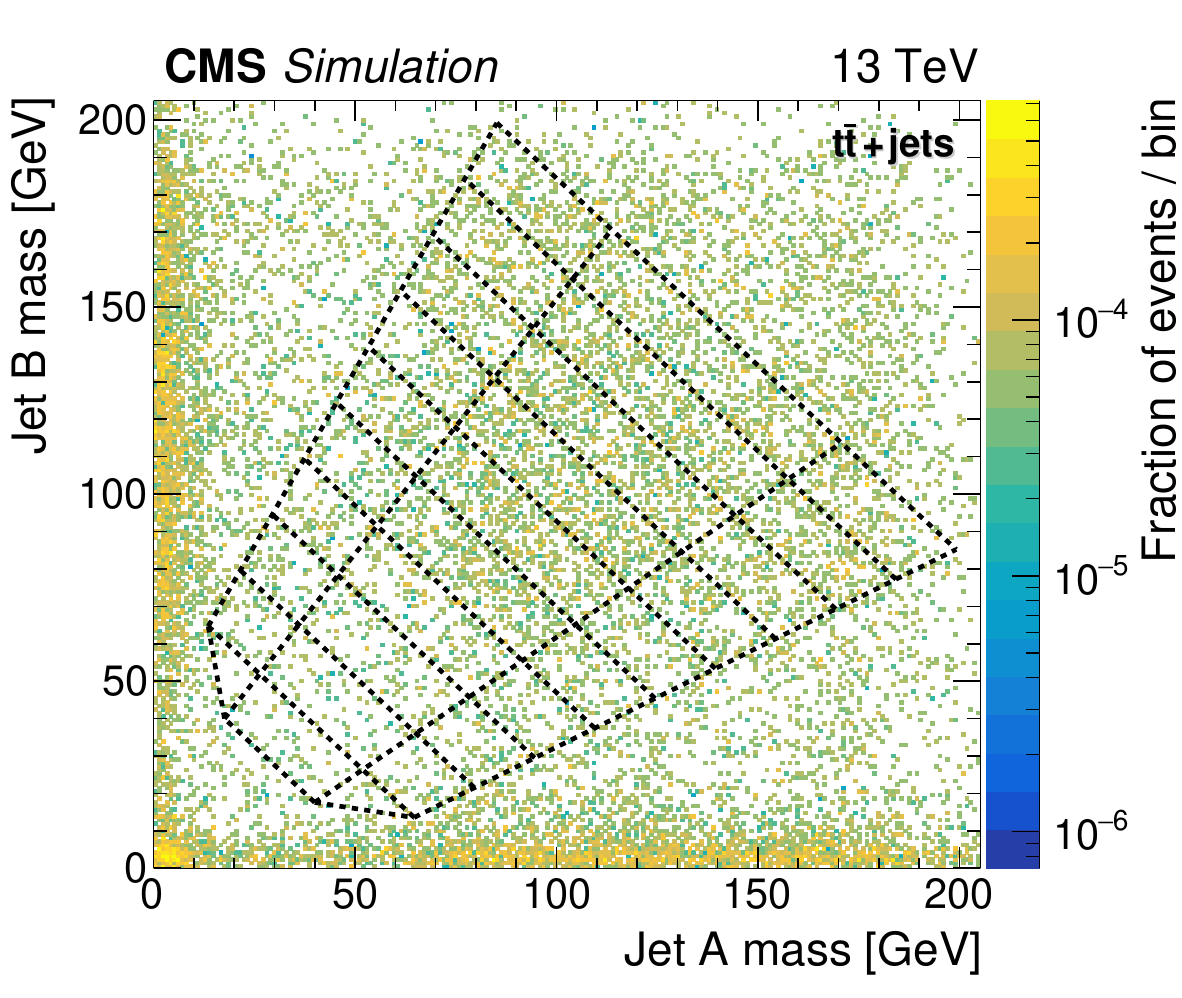}
\includegraphics[width=0.494\textwidth]{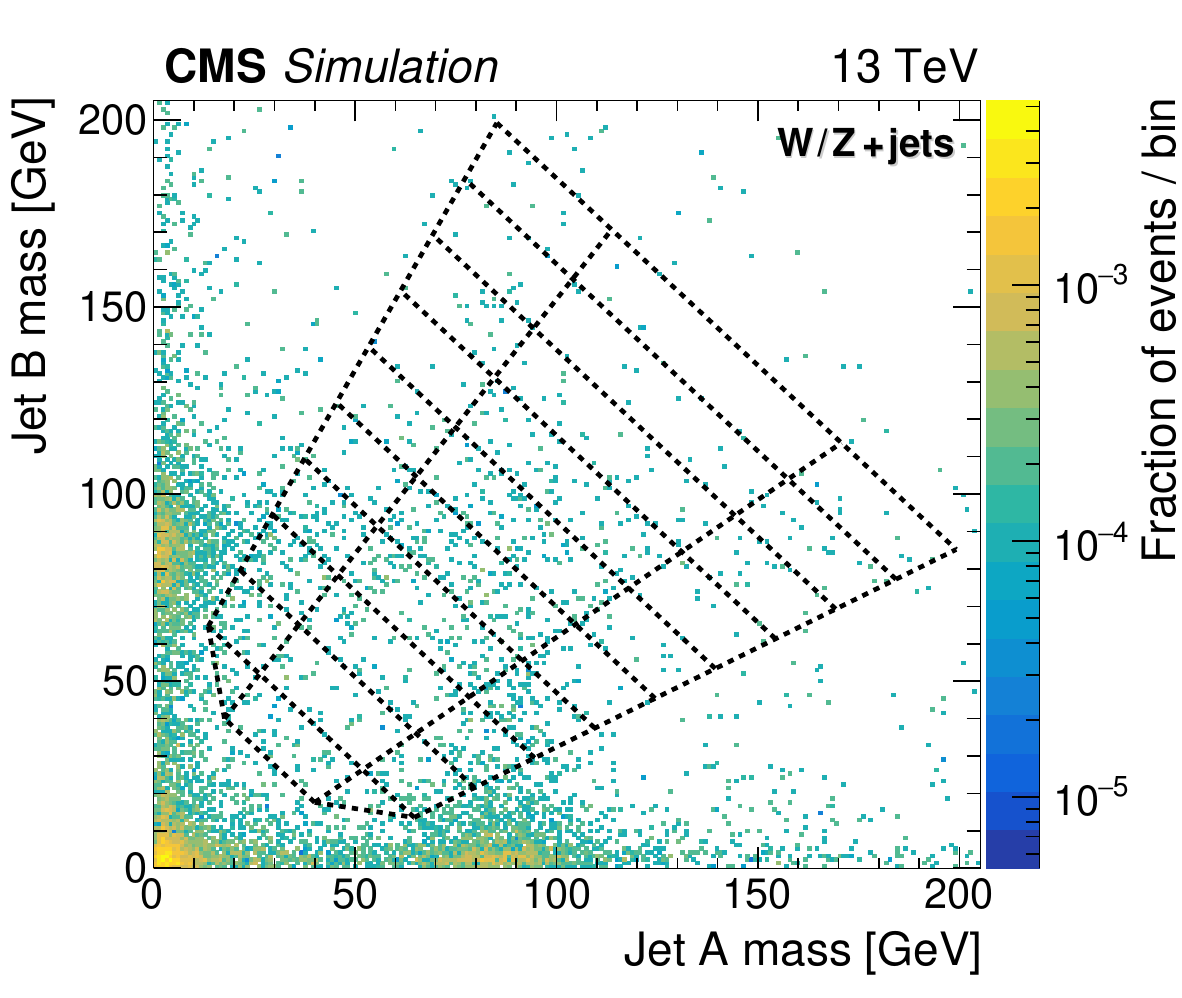}
\caption{
The normalised distribution of events in the 2D soft-drop mass plane
overlaid by the map of mass regions.
The upper left, upper right, and middle left panels correspond to signal events
for $\MSUSY = 2000\GeV$ and \MLH values of 40, 70, and 125\GeV, respectively.
The panels at middle right, lower left, and lower right correspond to simulated multijet, \ttbar, and vector boson backgrounds, respectively.
All events satisfy the \TR requirement and the kinematic selection.
}
\label{fig:signalMassPlots}
\end{figure*}

The event distributions
for a set of signal models with different \MLH values are shown in \fig~\ref{fig:signalMassPlots},
with the signal and sideband mass regions overlaid.
The peaks in the signal distributions where one or both AK8 jets have a soft-drop mass close to zero result from a selected jet originating from a single parton 
or one of the $\LH\to\bbbar$ decays lying outside the acceptance of the jet reconstruction algorithm. The latter can happen when the angular separation
of the \PQb~quarks exceeds the AK8 jet distance parameter, or when the ratio of the \PQb~quark \pt values is larger than 9 (such that the softer \PQb~quark would not satisfy the $z_{\textrm{cut}}$ threshold in the soft-drop algorithm).
For signal models with $40 < \MLH < 125\GeV$, $\approx\!50\%$ of the events that satisfy the kinematic and \TR selection fall within any of the $S_i$.
However, for $\MLH < 35\GeV$
the bulk of the distribution is lower in mass than $S_1$,
leading to a rapid decrease in signal acceptance.

{\tolerance=500
The distributions of background events are 
also shown in \fig~\ref{fig:signalMassPlots}.
The majority of multijet events contain at least one AK8 jet evaluated to have a small soft-drop mass,
reflecting the characteristic one-prong structure of quark and gluon jets.
After applying the kinematic and \TR selection criteria, approximately 5\% of multijet events fall within any of the $S_i$,
with greater probability at small masses.
For the vector boson and \ttbar backgrounds
the corresponding figures are 7 and 19\%, respectively,
concentrated in the $S_i$ corresponding to masses between the \PW~boson and top quark masses.
\par}

For each $S_i$ there are two corresponding sideband regions, \SB, 
used for the multijet background estimation 
 described in Section~\ref{sec_qcdEstimation}.
The sideband regions \SBo have a triangular form to avoid the
region of very small soft-drop masses, where the density from multijet events increases sharply.

\subsection{\texorpdfstring{Categorisation in \HT and expected yields}{Categorisation in HT and expected yields}}\label{sec_simulatedResults}

\begin{figure}[hbtp]
\centering
\includegraphics[width=\cmsFigWidthE]{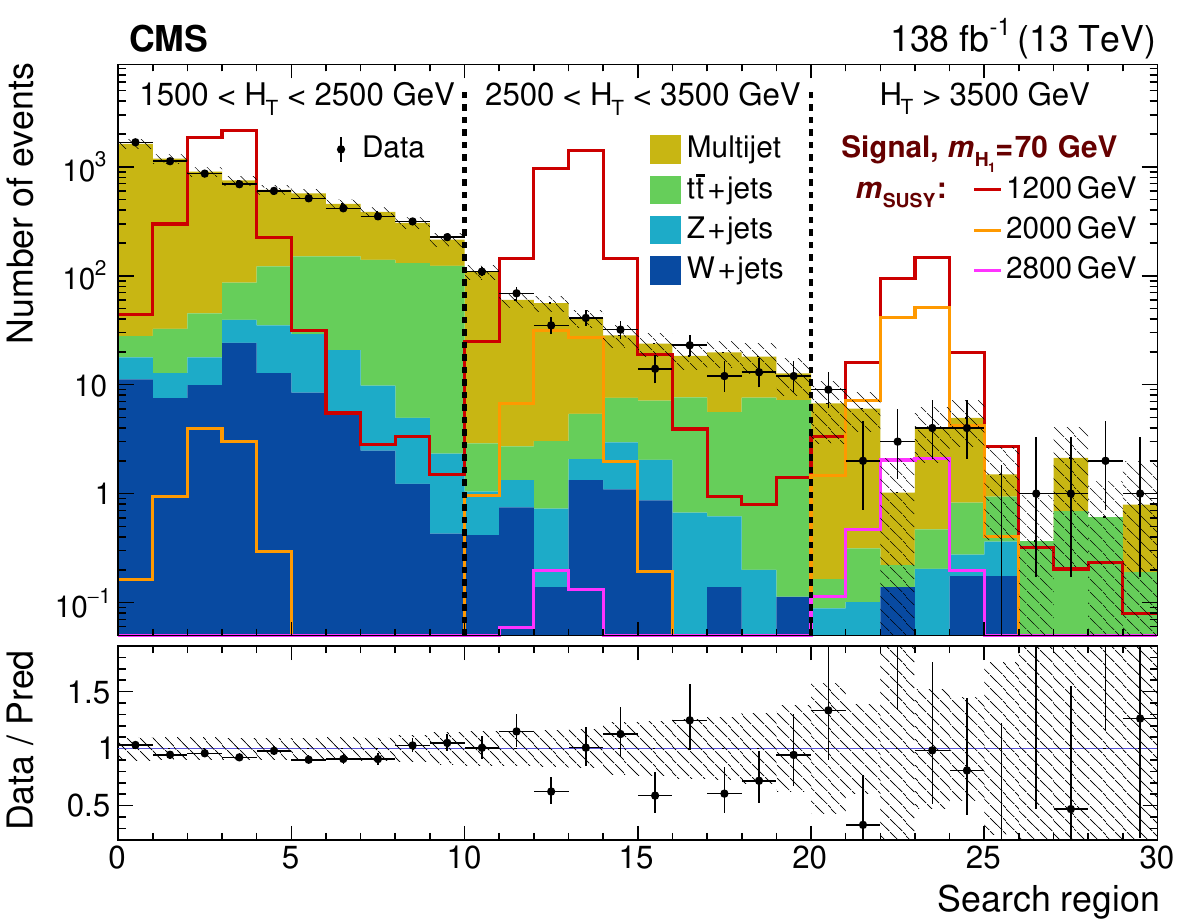}
\caption{
Observed and expected yields in the \TR for each of the 30 search regions, summed over the three data-taking years.
The multijet background is 
estimated from data using the method described in Section~\ref{sec_qcdEstimation},
while the other backgrounds are simulated.
Example signal distributions are shown for $\MLH = 70\GeV$ and $\MSUSY=1200$, 2000, and 2800\GeV.
The error bars represent the statistical uncertainties and the hatched bands the systematic uncertainties.
}
\label{fig:simResults}
\end{figure}

The selected events are classified according to three \HT categories: 1500--2500, 2500--3500, and above $3500\GeV$.
Each \HT category is divided into the 10 mass signal regions $S_i$ defined in \fig~\ref{fig:massGridEmpty},
resulting in a total of 30 search regions for each data-taking year.
As can be seen in \fig~\ref{fig:simResults} for \TR data summed over the three data-taking years, the search region yields can be visualised through a 30-bin histogram
where bins 1--10 represent the $S_i$, in ascending order,
for the first \HT category.
The subsequent two sets of 10 bins represent the results for the second
and third
\HT categories.
The primary background is from multijet events, estimated from data using the method described in Section~\ref{sec_qcdEstimation}.
The expected contribution from \ttbar events is also significant,
particularly in the larger soft-drop mass regions populated by
jets 
from
hadronic top quark or \PW boson decays.
The \ttbar simulation is validated in a dedicated \ttbar-enriched control region in data.
In \fig~\ref{fig:massGridEmpty} this is the triangular region of the parameter space with both jet masses below 200\GeV and above the upper boundary of mass region 10.
The yields from \Zjets and \Wjets production are small in comparison.
All expected SM backgrounds tend to exhibit small values of \HT compared to signal.

The distributions in signal events for
$\MLH = 70\GeV$ and $\MSUSY=1200$, 2000 and 2800\GeV
are also shown in \fig~\ref{fig:simResults}.
Although the production cross section decreases quickly with increasing \MSUSY,
the fraction of events in the larger-\HT categories increases.
Within each \HT category, the distribution of events in the 10 $S_i$ bins is described by a peak with a width of about three bins, centred near the model value of \MLH.

\section{Multijet background estimation from data}\label{sec_qcdEstimation}

The mass sideband regions \SB form a basis for using data to estimate the multijet background.
The density of the multijet background is approximately uniform within each of the 10 mass regions (spanning $S_i$ and \SB for each region $i$ illustrated in \fig~\ref{fig:massGridEmpty}).
Apart from \SBo, each \SB is constructed to have the same area as $S_i$
such that the corresponding multijet yields, respectively denoted \SBY and $\hat{S}_{i}$, are approximately equal.
The observed ratios of $S_i$ to \SB yields, $F_{i}$, are measured in \CR data.
The $F_{i}$ factors are found to be close to unity except for the $F_{1}$ values which are approximately 1.5.

The multijet background in the \TR is estimated independently for each signal region $S_i$:
\begin{linenomath} 
\begin{equation}
\label{abcdTag}
\hat{S}_{i}^{\textrm{\TR}} = F_{i} \, \SBY^{\textrm{\TR}},
\end{equation}
\end{linenomath} 
where $\SBY^{\textrm{\TR}}$ is the observed \TR yield in sideband region \SB after subtracting the contributions from the other simulated backgrounds.
In rare cases where the prediction $\hat{S}_{i}^{\textrm{\TR}}$ is negative, it is set equal to zero.

Since the $F_i$ factors are measured and applied in different regions of double-\PQb~tag discriminant space,
any correlation 
between the soft-drop mass and the double-\PQb~tag discriminant of AK8 jets can bias the prediction of \eq~(\ref{abcdTag}).
Using a sample of data satisfying an alternative kinematic event selection with the requirement for one or more AK4 jets inverted,
the variation of $F_i$ between the \TR and the \CR is found to be less than 10\%.

\begin{figure}[hbtp]
\centering
\includegraphics[width=\cmsFigWidthE]{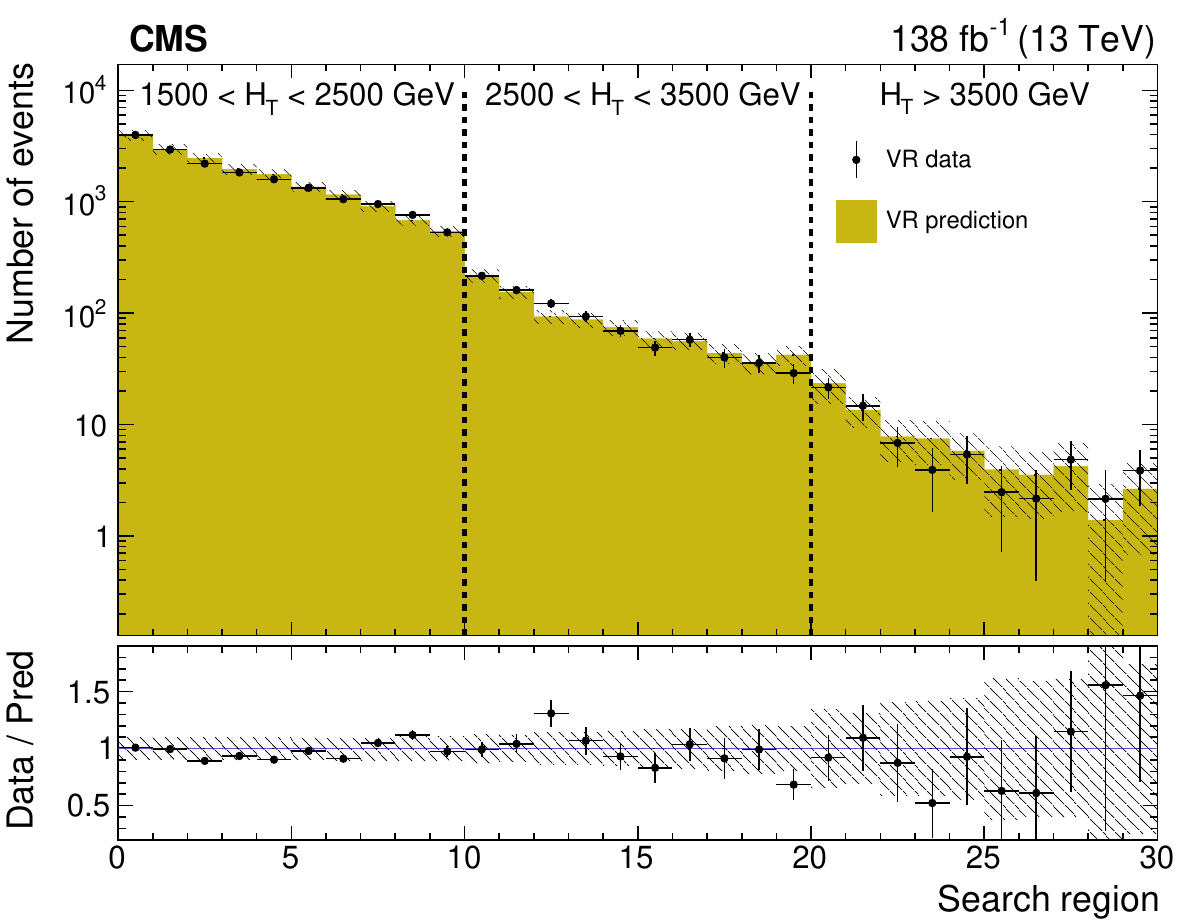}
\caption{
A comparison of the predicted and observed multijet yields in the validation region (\VR), after subtraction of the other simulated backgrounds.
The prediction is made separately for the three data-taking years,
and the results are summed.
The error bars on the data points represent their statistical uncertainties.
The uncertainties in the predicted yields (statistical and systematic) are indicated by the hatched bands.
}
\label{fig:histoControl}
\end{figure}

The overall accuracy of the multijet estimation is assessed through closure tests.
First the method is applied to simulated multijet events in the \TR
where, within statistical uncertainties, 
the predicted yields are consistent with the simulated yields for each data-taking year.
Second the method is applied in the multijet-dominated \VR data (defined in \fig~\ref{fig:dbtTagAntiControl}) by making the appropriate modification to \eq~(\ref{abcdTag}): $\hat{S}_{i}^{\textrm{\VR}} = F_i \, \SBY^{\textrm{\VR}}$.
The resulting predicted and observed \VR yields
are consistent within uncertainties, as shown in \fig~\ref{fig:histoControl}.
Based on the results of the closure tests, a systematic uncertainty of 15 (30\%) is assigned in the lower two \HT categories (upper \HT category).

\section{Systematic uncertainties}\label{sec_systematicUncertainties}

The simulated events for signal and the \ttbar, \Zjets, and \Wjets backgrounds are affected by various systematic uncertainties.
The efficiency for tagging (mistagging) a jet originating from two \PQb quarks (a light-flavour quark or gluon) is corrected to match that observed in data~\cite{BTV-16-002}. The uncertainty in this correction corresponds to $\approx\!10\%$
in the simulated signal and background yields.
The uncertainties related to the jet energy corrections are applied to the jet properties in bins of \pt and~$\eta$.
These uncertainties affect the event \HT,
leading to an $\approx\!4\%$ migration of events between adjacent \HT categories.
The uncertainty in the soft-drop mass scale in simulation relative to data
leads to a migration of events between adjacent $S_i$ and \SB regions of up to 10\%.
The uncertainty in the simulated soft-drop mass resolution
affects the widths of the simulated
mass peaks.
This effect is larger for signal models with small \MLH and can reduce the $S_i$ selection efficiency by up to 20\%.

The systematic uncertainties are assumed to be fully correlated among the data-taking years 
except for 
the 2016 double-\PQb tagging uncertainties, which are assumed uncorrelated because
the CMS pixel detector was upgraded prior to 2017 data-taking.
Changing these correlation assumptions is found to have only a small effect on the final results.
Systematic uncertainties related to integrated luminosity,
pileup, PDFs, renormalisation and factorisation scales, modelling of initial-state radiation, and background cross sections were also evaluated,
along with the statistical uncertainties in the simulation,
and were found to make negligible contributions to the total uncertainty.

Systematic uncertainties in multijet yields arise from the systematic uncertainties in the $F_i$ factors.
As described in Section~\ref{sec_qcdEstimation},
an uncertainty of $15\%$ is applied to the $F_i$ in the lower two \HT categories and $30\%$ in the upper \HT category,
uncorrelated among different $F_i$.
Except in the lowest \HT category, the total uncertainty in the multijet yield is dominated by the statistical uncertainty in 
$\SBY^{\textrm{\TR}}$.

\section{Results}\label{sec_results}

\begin{figure}[hbtp]
\centering
\includegraphics[width=\cmsFigWidthE]{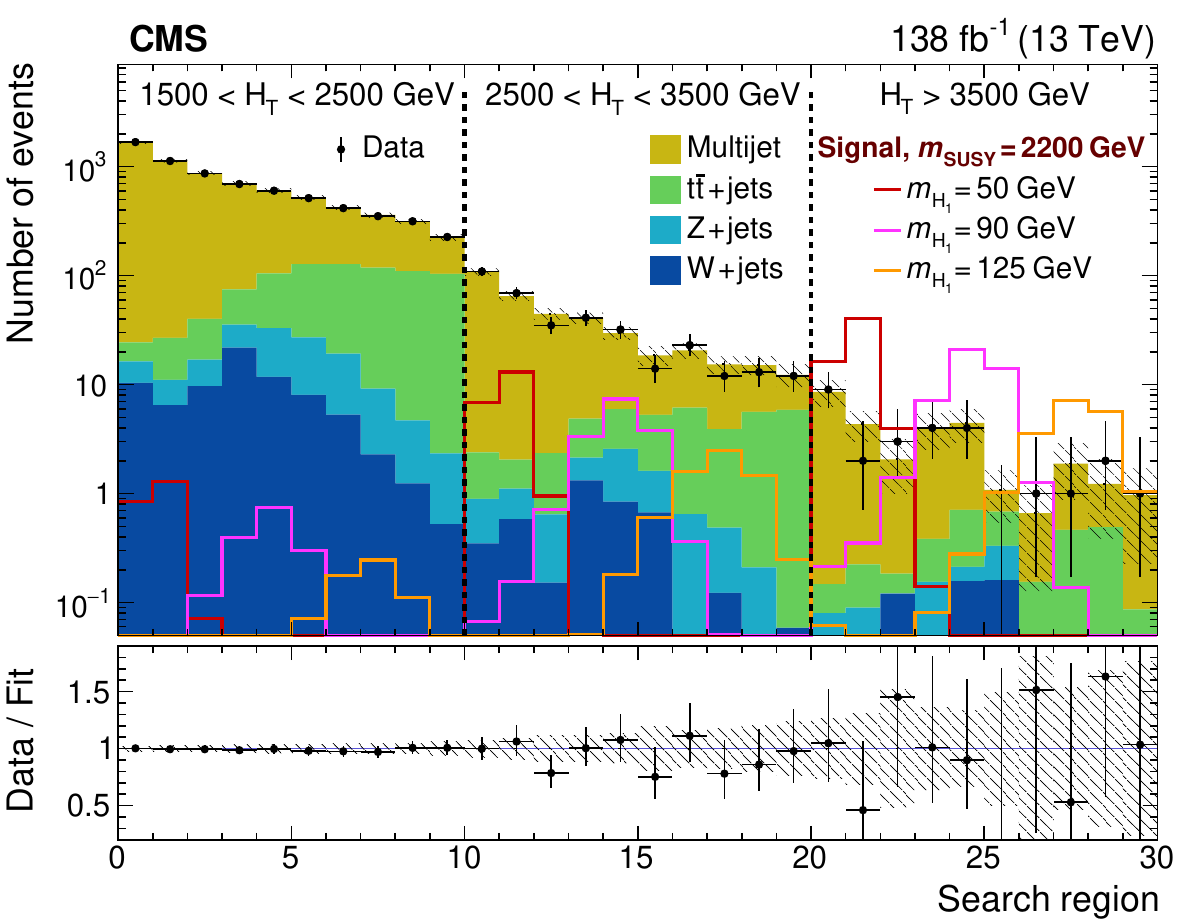}

\caption{
Yields in all search regions after the background-only fit, summed over the three data-taking years.
Example signal contributions used in the signal+background fits
are shown for $\MSUSY=2200\GeV$, and $\MLH=50$, 90, and 125\GeV.
The error bars represent the statistical uncertainties and the hatched bands the systematic uncertainties.
}
\label{fig:money}
\end{figure}

Binned maximum likelihood fits to the data in all 30 search regions $S_i$ for each data-taking year are carried out under background-only and signal+background hypotheses.
The corresponding sideband regions \SB are fitted simultaneously, thereby constraining the multijet contributions to the search region yields through \eq~(\ref{abcdTag}).
The likelihood functions are defined through the product of $90 \times 2$ Poisson distributions~\cite{ATLAS:2011tau}, one for each search region and one for each sideband region, with additional constraint terms for the ``nuisance'' parameters that account for the systematic uncertainties summarised in Section~\ref{sec_systematicUncertainties}.
\figfull~\ref{fig:money} compares the result of the background-only fit to the yields in the search regions for the combination of 2016, 2017, and 2018 data.
There is no evidence for deviations of the data from the fitted background.
The values and uncertainties of most nuisance parameters are unchanged in the fit, but the ones corresponding to the $F_i$ are constrained through \eq~(\ref{abcdTag}) when the yields $\hat{S}_{i}^{\textrm{\TR}}$ and $\SBY^{\textrm{\TR}}$ are sufficiently large.

\begin{figure}[hbtp]
\centering
\includegraphics[width=\cmsFigWidthC]{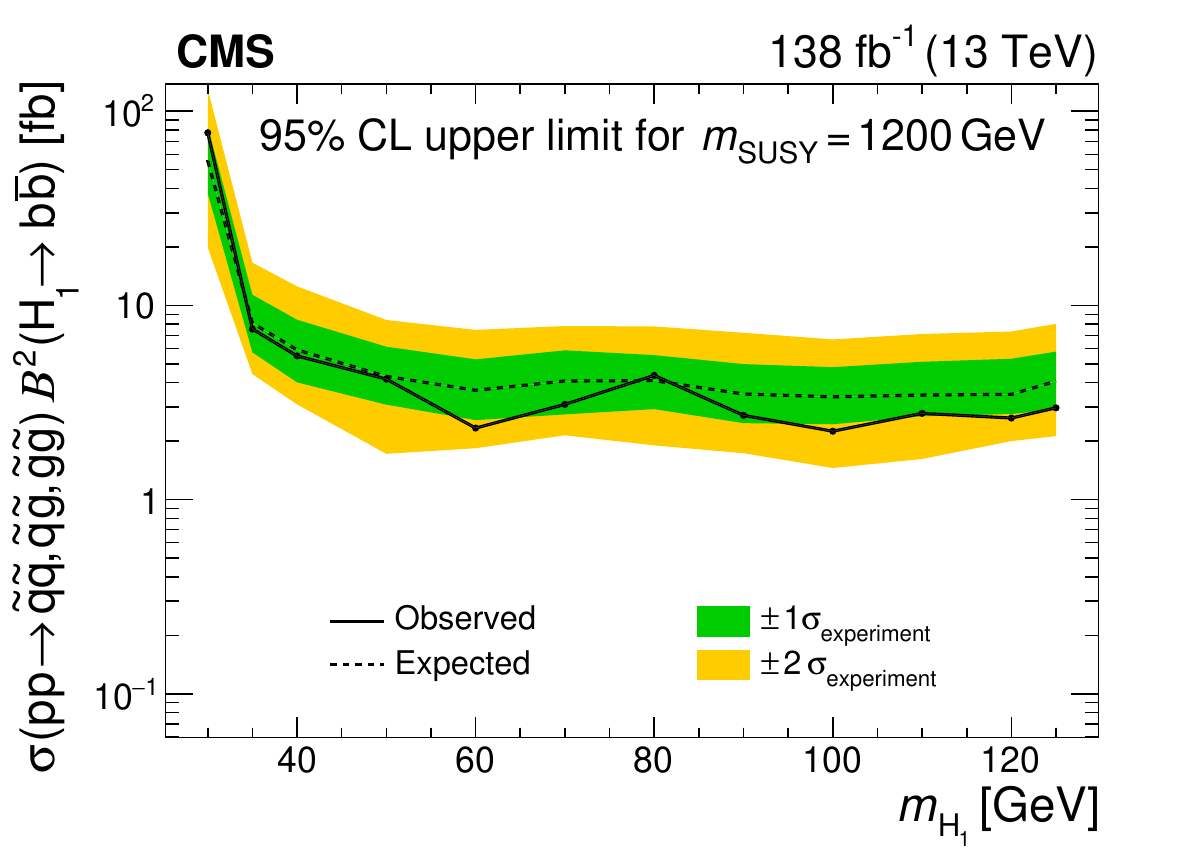}
\includegraphics[width=\cmsFigWidthC]{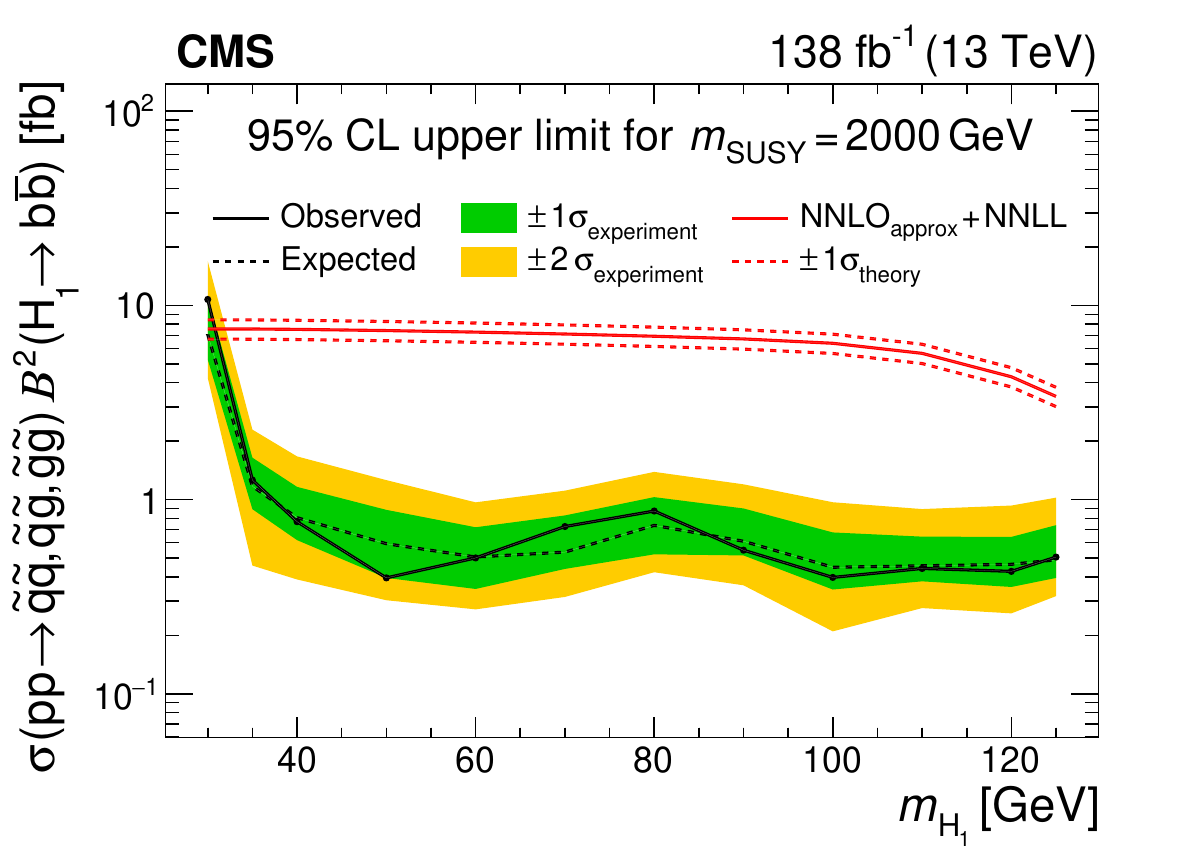}
\includegraphics[width=\cmsFigWidthC]{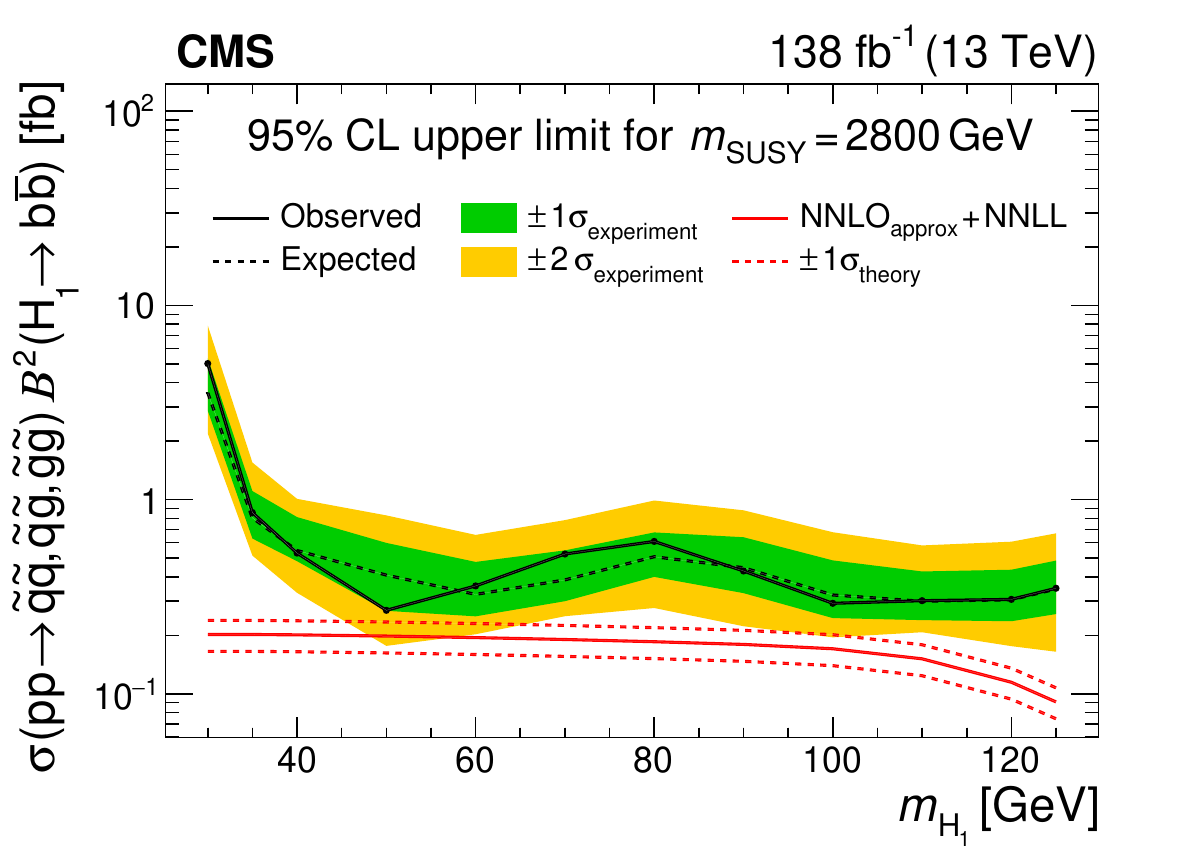}
\caption{
Upper limits at 95\% \CL on $\sigma\BR^2$ as a function of \MLH,
for \MSUSY values of 1200 (upper), 2000 (middle), and 2800\GeV (lower).
The solid and dashed black lines indicate the observed and median expected limits, respectively.
The inner (green) band and the outer (yellow) band indicate the regions containing 68 and 95\%, respectively, of the distribution of limits expected under the background-only hypothesis.
The solid and dashed red lines show the theoretical value of $\sigma\BR^2$ and its uncertainty~\cite{bib-nlo-nll-01,bib-nlo-nll-02,bib-nlo-nll-03,bib-nlo-nll-04,bib-nlo-nll-05,bib-nnll-02,bib-nnll-03,bib-nnll-04,bib-nnll,DJOUADI199856}.
In the upper plot, these $\sigma\BR^2$ values are beyond the maximum of the vertical axis.
}
\label{fig:obsLimitWithGluino_1D}
\end{figure}

Signal+background fits are used to set 95\% confidence level (\CL) upper limits on the product $\sigma\BR^2$
for the mass points in the benchmark signal model.
The limits are set using the modified frequentist \CLs criterion~\cite{Junk:1999kv,Read:2002hq},
with the profile likelihood ratio as test statistic~\cite{ATLAS:2011tau}.
The observed and expected 95\% \CL upper limits on $\sigma\BR^2$ are shown in \fig~\ref{fig:obsLimitWithGluino_1D},
as functions of \MLH for constant \MSUSY.
The upper limits are weaker for models with $\MLH<35\GeV$,
for which the signal-event distribution in the 2D soft-drop mass plane peaks outside the signal regions.
The limits 
have no significant dependence on \MSUSY
for models with $\MSUSY>2000\GeV$, whose signal events mostly populate the upper \HT category (as shown in \fig~\ref{fig:simResults}).

\begin{figure}[hbtp]
\centering
\includegraphics[width=\cmsFigWidthF]{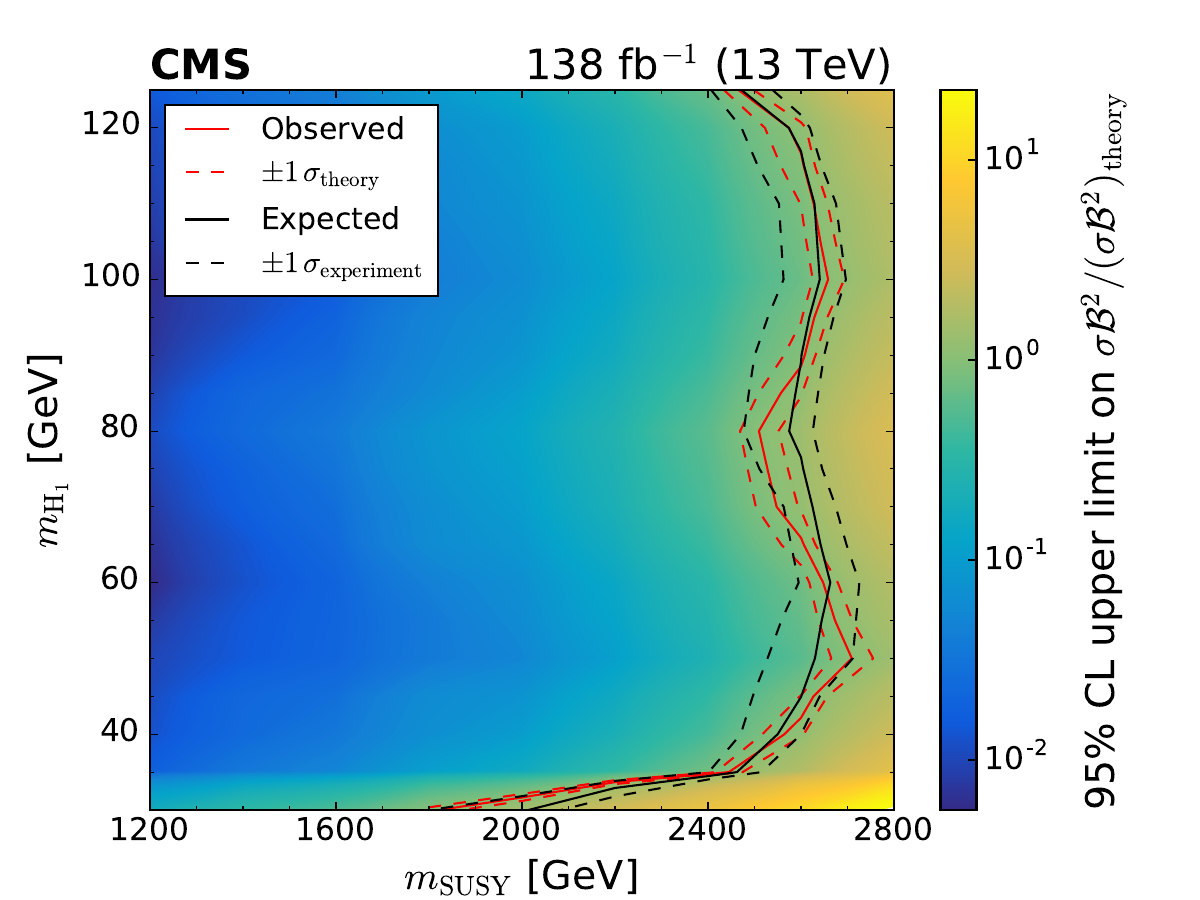}
\caption{
The observed 95\% \CL upper limit on $\sigma\BR^2 / (\sigma\BR^2)_{\textrm{theory}}$, quantified by the colour scale
as a function of \MLH and \MSUSY.
The solid and dashed red lines indicate the observed excluded region and its theoretical uncertainty, respectively.
The solid and dashed black lines respectively represent the expected excluded region and its 68\% \CL interval, under the background-only hypothesis.
}
\label{fig:obsLimitWithGluino}
\end{figure}

The $\sigma\BR^2$ upper limits are used in conjunction
with the theoretical $\sigma$ and \BR values from Section~\ref{sec_signalModelAndMotivation}
to exclude ranges of masses in \MLH and \MSUSY in the benchmark model.
The observed 95\% \CL upper limits on $r$, the ratio of measured and theoretical values of $\sigma\BR^2$, are shown in \fig~\ref{fig:obsLimitWithGluino}, with the corresponding exclusion contours at $r=1$.
Masses $1200 < \MSUSY < 2500\GeV$ are excluded within the range $40 < \MLH < 120\GeV$.
Expected exclusion contours for the background-only scenario agree within one standard deviation with the observed contours.
In the region $110 < \MLH < 125\GeV$, \BR starts to decrease more quickly (as shown in Table~\ref{table:sigSamples_MH}),
leading to a corresponding reduction in sensitivity.
Most of the sensitivity at large \MSUSY comes from the $\HT > 3500\GeV$ region,
where
the statistical uncertainties in the observed yields are dominant over systematic uncertainties.
This search does not explore the region outside of that shown in \fig~\ref{fig:obsLimitWithGluino}.

To aid reinterpretation of the search by reducing the model-dependence, limits evaluated using only the upper \HT category are presented in Appendix~\ref{app_A}.
Tabulated results are provided in the HEPData record for this analysis~\cite{hepdata}.

\ifthenelse{\boolean{cms@external}}{}{\pagebreak}
\section{Summary}\label{sec_summary}

This paper presents a search for pairs of light Higgs bosons (\LH) produced in supersymmetric cascade decays.
The targeted final states have small amounts of missing transverse momentum and two $\LH\to\bbbar$ decays
that are reconstructed as large-radius jets using substructure techniques.
The search is based on data from $\Pp\Pp$ collisions
collected by the CMS experiment at $\sqrt{s}=13\TeV$ during 2016--2018,
corresponding to an integrated luminosity of 138\fbinv.

With no evidence found for an excess of events beyond the background expectations of the standard model (SM),
the results are interpreted
in the next-to-minimal supersymmetric extension of the SM (NMSSM),
where a ``singlino'' of small mass leads to squark and gluino cascade decays that can predominantly end in a highly Lorentz-boosted singlet-like \LH
and a singlino-like neutralino of small transverse momentum.

Upper limits are set on the
product of the production cross section and the square of the $\bbbar$ branching fraction of the \LH
for an NMSSM benchmark model with almost mass-degenerate gluinos and light-flavour squarks
and branching fractions of unity for the cascade decays ending with the \LH.
Under the assumption of an SM-like $\LH\to\bbbar$ branching fraction, 
\LH bosons with masses in the range 40--120\GeV
arising from the decays of squarks or gluinos with a mass of
1200 to 2500\GeV
are excluded at 95\% confidence level.

\begin{acknowledgments}
We congratulate our colleagues in the CERN accelerator departments for the excellent performance of the LHC and thank the technical and administrative staffs at CERN and at other CMS institutes for their contributions to the success of the CMS effort. In addition, we gratefully acknowledge the computing centres and personnel of the Worldwide LHC Computing Grid and other centres for delivering so effectively the computing infrastructure essential to our analyses. Finally, we acknowledge the enduring support for the construction and operation of the LHC, the CMS detector, and the supporting computing infrastructure provided by the following funding agencies: BMBWF and FWF (Austria); FNRS and FWO (Belgium); CNPq, CAPES, FAPERJ, FAPERGS, and FAPESP (Brazil); MES and BNSF (Bulgaria); CERN; CAS, MoST, and NSFC (China); MINCIENCIAS (Colombia); MSES and CSF (Croatia); RIF (Cyprus); SENESCYT (Ecuador); MoER, ERC PUT and ERDF (Estonia); Academy of Finland, MEC, and HIP (Finland); CEA and CNRS/IN2P3 (France); BMBF, DFG, and HGF (Germany); GSRI (Greece); NKFIH (Hungary); DAE and DST (India); IPM (Iran); SFI (Ireland); INFN (Italy); MSIP and NRF (Republic of Korea); MES (Latvia); LAS (Lithuania); MOE and UM (Malaysia); BUAP, CINVESTAV, CONACYT, LNS, SEP, and UASLP-FAI (Mexico); MOS (Montenegro); MBIE (New Zealand); PAEC (Pakistan); MES and NSC (Poland); FCT (Portugal); MESTD (Serbia); MCIN/AEI and PCTI (Spain); MOSTR (Sri Lanka); Swiss Funding Agencies (Switzerland); MST (Taipei); MHESI and NSTDA (Thailand); TUBITAK and TENMAK (Turkey); NASU (Ukraine); STFC (United Kingdom); DOE and NSF (USA).

\hyphenation{Rachada-pisek} Individuals have received support from the Marie-Curie programme and the European Research Council and Horizon 2020 Grant, contract Nos.\ 675440, 724704, 752730, 758316, 765710, 824093, 884104, and COST Action CA16108 (European Union); the Leventis Foundation; the Alfred P.\ Sloan Foundation; the Alexander von Humboldt Foundation; the Belgian Federal Science Policy Office; the Fonds pour la Formation \`a la Recherche dans l'Industrie et dans l'Agriculture (FRIA-Belgium); the Agentschap voor Innovatie door Wetenschap en Technologie (IWT-Belgium); the F.R.S.-FNRS and FWO (Belgium) under the ``Excellence of Science -- EOS" -- be.h project n.\ 30820817; the Beijing Municipal Science \& Technology Commission, No. Z191100007219010; the Ministry of Education, Youth and Sports (MEYS) of the Czech Republic; the Hellenic Foundation for Research and Innovation (HFRI), Project Number 2288 (Greece); the Deutsche Forschungsgemeinschaft (DFG), under Germany's Excellence Strategy -- EXC 2121 ``Quantum Universe" -- 390833306, and under project number 400140256 - GRK2497; the Hungarian Academy of Sciences, the New National Excellence Program - \'UNKP, the NKFIH research grants K 124845, K 124850, K 128713, K 128786, K 129058, K 131991, K 133046, K 138136, K 143460, K 143477, 2020-2.2.1-ED-2021-00181, and TKP2021-NKTA-64 (Hungary); the Council of Science and Industrial Research, India; the Latvian Council of Science; the Ministry of Education and Science, project no. 2022/WK/14, and the National Science Center, contracts Opus 2021/41/B/ST2/01369 and 2021/43/B/ST2/01552 (Poland); the Funda\c{c}\~ao para a Ci\^encia e a Tecnologia, grant CEECIND/01334/2018 (Portugal); the National Priorities Research Program by Qatar National Research Fund; MCIN/AEI/10.13039/501100011033, ERDF ``a way of making Europe", and the Programa Estatal de Fomento de la Investigaci{\'o}n Cient{\'i}fica y T{\'e}cnica de Excelencia Mar\'{\i}a de Maeztu, grant MDM-2017-0765 and Programa Severo Ochoa del Principado de Asturias (Spain); the Chulalongkorn Academic into Its 2nd Century Project Advancement Project, and the National Science, Research and Innovation Fund via the Program Management Unit for Human Resources \& Institutional Development, Research and Innovation, grant B05F650021 (Thailand); the Kavli Foundation; the Nvidia Corporation; the SuperMicro Corporation; the Welch Foundation, contract C-1845; and the Weston Havens Foundation (USA).
\end{acknowledgments}
\bibliography{auto_generated}

\clearpage
\numberwithin{table}{section}
\numberwithin{figure}{section}
\appendix
\section{Simplified analysis for reinterpretation}\label{app_A}

\begin{figure}[hbtp]
    \centering
    \includegraphics[width=\cmsFigWidthC]{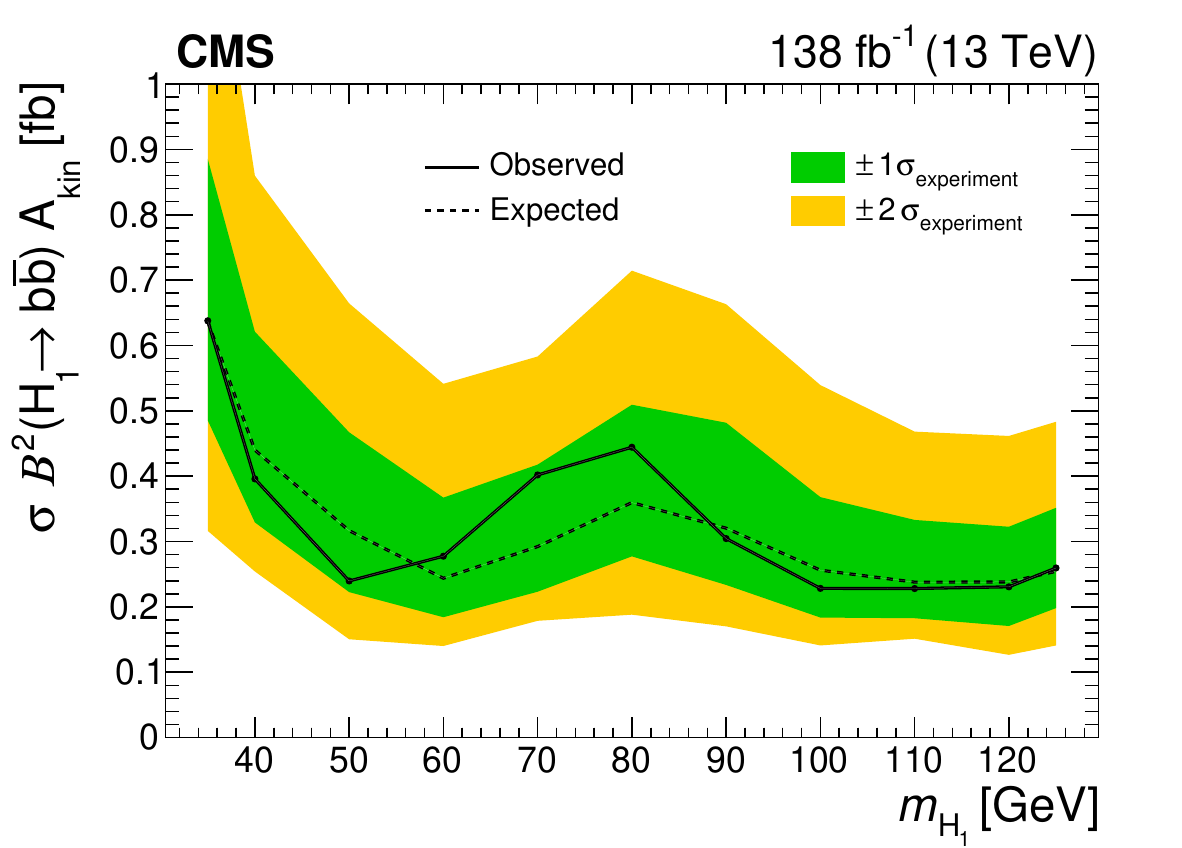}
    \caption{
    The observed and expected 95\% \CL upper limit on the product of $\sigma\BR^2$ and \Akin, the kinematic acceptance and efficiency for the $\HT>3500\GeV$ region, as a function of \MLH.
    The results are independent of \MSUSY within 10\% in the range $1600<\MSUSY<2800\GeV$.
    The solid and dashed black lines indicate the observed and median expected limits, respectively.
    The inner (green) band and the outer (yellow) band indicate the regions containing 68 and 95\%, respectively, of the distribution of limits expected under the background-only hypothesis.
    }
    \label{fig:obsLimitSimplified}
    \end{figure}

To aid reinterpretation of the search, a simplified analysis is performed using only the 10 search regions in the upper \HT category.
The value \Akin is defined as the product of acceptance and efficiency for a signal event to satisfy the kinematic selection (defined in Section~\ref{sec_eventSelection}) and the $\HT>3500\GeV$ requirement. 
The value of \Akin is 
common among all
10 search regions in the simplified analysis, and is quoted for the benchmark signal model in Table~\ref{table:kinAccEff}.
Upper limits on the product $\sigma \, \BR^2 \, \Akin$ as a function of \MLH are set in \fig~\ref{fig:obsLimitSimplified}, from which $\sigma\BR^2$ limits for different signal models can be derived through division by the appropriate value of \Akin.
Since the upper \HT category provides most of the sensitivity for $\MSUSY>2000\GeV$ in the nominal analysis, 
the $\sigma\BR^2$ upper limits in this region are not much weaker in the simplified analysis.
This is not the case in the region $\MSUSY<2000\GeV$, where the lower \HT categories become important.

\begin{table}[hbtp]
  \centering
  \topcaption{
  Reference values of the product of kinematic acceptance and efficiency (\Akin) for the $\HT>3500\GeV$ region for the benchmark signal model with different values of \MSUSY.
  These values are independent of \MLH within 2\% in the range $30<\MLH<125\GeV$.
  }
  \begin{tabular}{c|cccccc}
    \MSUSY [{\GeVns}] & 1600 & 2000 & 2200 & 2400 & 2600 & 2800 \\ [0.3ex]
    \hline
    \Trule
    \Akin & 0.17 & 0.46 & 0.58 & 0.66 & 0.71 & 0.74 \\
  \end{tabular}
  \label{table:kinAccEff}
  \end{table}

The double-\PQb~tag and mass region selections are not considered in \Akin. This is done for simplicity, and because the fraction of events satisfying these selections is not found to be strongly model-dependent (except for the dependence on \MLH, which is accounted for explicitly in \fig~\ref{fig:obsLimitSimplified}).
For the benchmark model, this fraction is found to be independent of \MSUSY within 10\% in the region $1600<\MSUSY<2800\GeV$ and $35<\MLH<125\GeV$.
This approximate independence does not hold for models with $\MSUSY<1600\GeV$, 
where the \LH \pt distribution has substantial contributions below the \pt necessary for the $\LH\to\bbbar$ decay products to be merged in a single AK8 jet.
Only models with typical $\bbbar$ angular separation $\Delta R<0.8$ should be considered for reinterpretation.
\cleardoublepage \section{The CMS Collaboration \label{app:collab}}\begin{sloppypar}\hyphenpenalty=5000\widowpenalty=500\clubpenalty=5000
\cmsinstitute{Yerevan Physics Institute, Yerevan, Armenia}
{\tolerance=6000
A.~Tumasyan\cmsAuthorMark{1}\cmsorcid{0009-0000-0684-6742}
\par}
\cmsinstitute{Institut f\"{u}r Hochenergiephysik, Vienna, Austria}
{\tolerance=6000
W.~Adam\cmsorcid{0000-0001-9099-4341}, J.W.~Andrejkovic, T.~Bergauer\cmsorcid{0000-0002-5786-0293}, S.~Chatterjee\cmsorcid{0000-0003-2660-0349}, K.~Damanakis\cmsorcid{0000-0001-5389-2872}, M.~Dragicevic\cmsorcid{0000-0003-1967-6783}, A.~Escalante~Del~Valle\cmsorcid{0000-0002-9702-6359}, R.~Fr\"{u}hwirth\cmsAuthorMark{2}\cmsorcid{0000-0002-0054-3369}, M.~Jeitler\cmsAuthorMark{2}\cmsorcid{0000-0002-5141-9560}, N.~Krammer\cmsorcid{0000-0002-0548-0985}, L.~Lechner\cmsorcid{0000-0002-3065-1141}, D.~Liko\cmsorcid{0000-0002-3380-473X}, I.~Mikulec\cmsorcid{0000-0003-0385-2746}, P.~Paulitsch, F.M.~Pitters, J.~Schieck\cmsAuthorMark{2}\cmsorcid{0000-0002-1058-8093}, R.~Sch\"{o}fbeck\cmsorcid{0000-0002-2332-8784}, D.~Schwarz\cmsorcid{0000-0002-3821-7331}, S.~Templ\cmsorcid{0000-0003-3137-5692}, W.~Waltenberger\cmsorcid{0000-0002-6215-7228}, C.-E.~Wulz\cmsAuthorMark{2}\cmsorcid{0000-0001-9226-5812}
% ...
\par}
\vskip\cmsinstskip
\dag:~Deceased\\
$^{1}$Also at Yerevan State University, Yerevan, Armenia\\
$^{2}$Also at TU Wien, Vienna, Austria\\
% ...
\end{sloppypar}
\end{document}